\documentclass[twocolumn, numberedappendix, twocolappendix,apj, iop, a4paper]{openjournal}

\usepackage{graphicx}
\usepackage{dcolumn}
\usepackage{bm}
\usepackage{physics}
\usepackage{color}
\usepackage{bbm} 
\definecolor{Blue}{rgb}{0,0.08,0.65}
\definecolor{Green}{rgb}{0.2,0.55,0.35}
\definecolor{grey}{rgb}{0.75,0.75,0.75}
\definecolor{Orange}{rgb}{1.0,0.5,0.15}
\definecolor{brown}{rgb}{0.7,0.25,0.0}
\definecolor{Pink}{rgb}{1.0,0.5,0.5}
\definecolor{darkerred}{rgb}{0.8,0,0}
\definecolor{darkerblue}{rgb}{0,0,0.8}
\definecolor{darkergreen}{rgb}{0,0.5,0}
\definecolor{darkcyan}{rgb}{0,0.6,0.6}

\usepackage[normalem]{ulem}
\usepackage{placeins} 
\usepackage{float} 
\usepackage{hyperref}  
\hypersetup{
colorlinks=true,
linkcolor=Blue,
filecolor=red,
citecolor=Blue,
urlcolor=Blue
}

\newcommand{\del}{\partial}

\newcommand{\kk}{\bm{k}}
\newcommand{\xx}{\bm{x}}

\newcommand{\be}{\begin{equation}}
\newcommand{\ee}{\end{equation}}
\newcommand{\bea}{\begin{eqnarray}}
\newcommand{\eea}{\end{eqnarray}}
\definecolor{darkgreen}{rgb}{0,0.5,0}

\begin{document}
\def\arraystretch{1.5}

\title{ When to interfere with dark matter? The impact of wave dynamics on statistics }


\author[0000-0002-1524-6949]{
 Alex Gough$^{\star,1}$}
\thanks{$^\star$\href{mailto:a.gough2@newcastle.ac.uk}
{a.gough2@newcastle.ac.uk}}
\author[0000-0001-7831-1579]{Cora Uhlemann$^{1,2}$}
\affiliation{$^{1}$School of Mathematics, Statistics and Physics, Newcastle University, Herschel Building, NE1 7RU Newcastle upon Tyne, U.K.}
\affiliation{$^2$ Fakult\"at f\"ur Physik, Universit\"at Bielefeld, Postfach 100131, 33501 Bielefeld, Germany}

\begin{abstract}

Ultralight candidates for dark matter can present wavelike features on astrophysical scales. Full wave based simulations of such candidates are currently limited to box sizes of 1--10 Mpc/$h$ on a side, limiting our understanding of the impact of wave dynamics on the scale of the cosmic web. We present a statistical analysis of density fields produced by perturbative forward models in boxes of 128 Mpc/$h$ side length. Our wave-based perturbation theory maintains interference on all scales, and is compared to fluid dynamics of Lagrangian perturbation theory. The impact of suppressed power in the initial conditions and interference effects caused by wave dynamics can then be disentangled. We find that changing the initial conditions captures most of the change in one-point statistics such as the skewness of the density field. However, different environments of the cosmic web, quantified by critical points of the  smoothed density, appear to be more sensitive to interference effects sourced by the quantum potential. This suggests that certain large-scale summary statistics may need additional care when studying cosmologies with wavelike dark matter.
\end{abstract}

\maketitle

\section{Introduction}

The current standard model of cosmology consists of cold dark matter and a cosmological constant. This $\Lambda$CDM model has been extremely successful at describing cosmological observations across a wide range of length and time scales \citep{ Frenk.White_2012_DarkMatter, Bull.etal_2016_LCDMProblems,PlanckCollaborationOverview}, despite the fundamental nature of dark matter and dark energy remaining unknown. Ongoing and upcoming cosmological surveys such as \textit{Euclid} \citep{Euclid_mission}, Rubin Observatory LSST \citep{LSST_mission}, and DESI \citep{DESI_mission} will provide an enormous amount of data which will allow precision measurement of cosmological parameters, and provide insight into the fundamental nature of these unknown components.

An alternative to the cold dark matter of $\Lambda$CDM which has gained a large amount of attention is dark matter which is fundamentally wavelike on astrophysical or cosmological scales. Such dark matter candidates were originally motivated out of particle physics \citep{Peccei1977PhRvL, Svrcek2006JHEP, Arvanitaki2010PhRvD, Hui2017, Jaeckel2022arXiv} though their phenomenology can also act to solve small scale challenges in $\Lambda$CDM \citep{Douspis:2019, DiValentino:2021, Perivolaropoulos:2021}. While a multitude of models exist which can encode self interactions or further degrees of freedom, here we concern ourselves with the simplest fundamentally wavelike dark matter candidate, a non-relativistic, ultralight scalar field, which we will refer to throughout as Fuzzy Dark Matter (FDM). Current constraints on various particle properties of axion models specifically, combining both particle and astrophysical constraints can be found in \citet{AxionLimits, O'Hare2024arXiv}. 

Fully non-linear descriptions of FDM are currently are limited to box sizes of $1$--$10$ Mpc at $z\simeq 3$ for boson masses of $m\sim 10^{-23} \ \mathrm{eV}/c^2$ \citep{MaySpringel2021, May.Springel_2023_HaloMass}. As such, investigations of the impact of FDM on larger cosmological scales require approximation techniques which neglect the full dynamics. Many of these techniques rely on modelling only the suppression of power on small scales which is present in the initial conditions of FDM, and neglect the interference effects and role of the quantum potential in the dynamics of the dark matter, an approached that has been dubbed ``classical FDM'' \citep{Dome.etal_2023_CosmicWebElongation, Dome.etal_2023_CosmicWebDissection}. As such, those interference effects are not fully represented in the final density distribution of the dark matter.

\begin{figure*}[p]
    \centering
    \includegraphics[width=\textwidth]{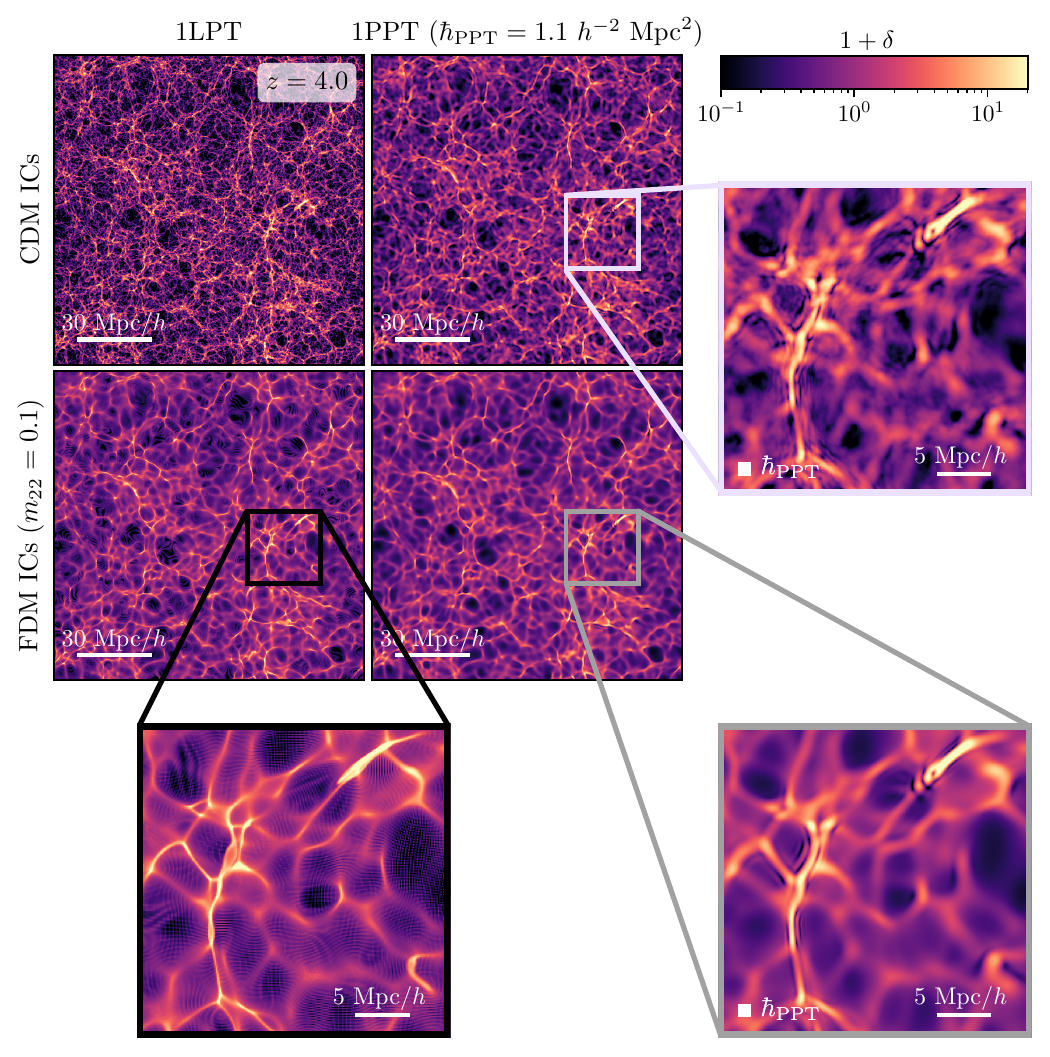}
    \caption[The density field for CDM and FDM initial conditions evolved with LPT and PPT.]{Projected $0.125 \ \mathrm{Mpc}/h$ thick slice of the dark matter density field at $z=4.0$. All plots are shown on the same logarithmic colour scale. These simulations have a box size of $128 \ \mathrm{Mpc}/h$ and a grid resolution of $0.125 \ \mathrm{Mpc}/h$. The LPT runs traced the positions of $(1024)^3$ particles. The zoomed inset regions are $30 \ \mathrm{Mpc}/h$ on a side. The size of the semiclassical parameter $\hbar_{\rm PPT}$ is shown as a white square in the zoomed plots. Columns in the main 2-by-2 correspond to which perturbation theory method is used: classical LPT fluid dynamics or wave PPT dynamics. The rows correspond to the initial conditions, either standard CDM initial conditions, or fuzzy dark matter initial conditions which suppress structure on small scales. The mass for the FDM initial conditions is chosen to be $10^{-23} \ \mathrm{eV}/c^2$. Notice that while classical evolution on FDM initial conditions produces a similar large scale cosmic web to the PPT+FDM ICs case, but by construction it cannot produce interference ripples seen in the wave evolution cases.}  
     \label{fig:slices-of-log-delta}
\end{figure*}

Figure~\ref{fig:slices-of-log-delta} shows a slice of the density field produced from the perturbative forward model used in this work. We can directly see that while classical FDM (the black outlined inset) does produce the overall smoothing seen in the full wave physics case (grey inset), by construction it cannot produce the interference fringes which occur due to the quantum potential. This work aims to investigate how justified such approaches are on large cosmological scales, separating the impact of dynamics and initial conditions on larger scales than can be probed with full numerical simulations. The fine grained pattern in the lower left plot is an artefact of the assignment of particles to the grid in LPT fields. This effect disappears when the density field is smoothed, and becomes less noticeable for higher order mass assignment kernels, as shown in Appendix \ref{app:sec:MAS_impact}. 

\subsection{Numerical Challenges for FDM}

In FDM, a single wavefunction $\psi$ describes the dark matter field of a particle with mass $m$. This wavefunction evolves according to the cosmological Schr\"odinger-Poisson equations  \citep[][]{WidrowKaiser1993,Guth.etal_2015_DarkMatter, Marsh_2016_AxionCosmology, Hui2021, Ferreira2021A&ARv, O'Hare2024arXiv}
\begin{subequations} \label{eqn:SP-full-eqns}
    \begin{align}
        i\hbar \partial_t \psi &= -\frac{\hbar^2}{2ma^2} \nabla^2 \psi + m V_N \psi\,, \\
        \nabla^2 V_N &= \frac{3 \Omega_m^0 H_0^2}{2}  \frac{\abs{\psi}^2 - 1}{a}.
    \end{align}
\end{subequations}
Here $V_N$ is the Newtonian gravitational potential, $\Omega_m^0$ is the current matter energy density and $H_0$ is the current Hubble parameter. The wavefunction described here is co-moving, meaning that the physical matter density would scale as $a^{-3/2}$ times this wavefunction. These equations arise as the non-relativistic limit of the Klein-Gordon equation of a scalar field in expanding space, see e.g. \cite{Marsh_2016_AxionCosmology} for a derivation.

The principle challenge in numerically solving the Schr\"odinger-Poisson equations \eqref{eqn:SP-full-eqns}
for the wavefunction while resolving wave interference scales are strict space and time resolution requirements which are dependent on the mass of the FDM particle. For pseudospectral methods such as those used in \citep{MaySpringel2021, May.Springel_2023_HaloMass} the time step requirement is
\begin{equation}
	\Delta t < \operatorname{min}\left(\frac{4}{3\pi} \frac{m}{\hbar} a^2 (\Delta x)^2, 2\pi \frac{\hbar}{m} \frac{1}{\abs{V_{\rm N, max}}}\right),
\end{equation}
where $\Delta x$ is the spatial resolution and $V_{\rm N, max}$ is the maximum of the gravitational potential. This is imposed to avoid aliasing in the wavefunction where the phase would increase by more than $2\pi$ in adjacent cells. Different numerical schemes will have different exact prescriptions for this form, but all have the scaling that $\Delta t \sim \Delta x^2$ owing to the diffusive nature of the Schr\"odinger equation. The FDM particle mass itself imposes a restriction on the spatial resolution through its de Broglie wavelength. If we wish to resolve velocities of size $v_{\rm max}$ in the simulation the spatial grid needs to resolve the de Broglie wavelength associated with that velocity
\begin{equation}
\Delta x < \frac{\hbar}{m}\frac{\pi}{v_{\rm max}}\,.
\end{equation}

These joint constraints on $\Delta x$ and $\Delta t$ are what make direct, large box, simulations of fuzzy dark matter so computationally expensive, as going to cosmological boxes $\order{100}$ Mpc/$h$ cannot be achieved by simply decreasing the spatial/temporal resolution.

\subsection{Summary of numerical approaches}

A variety of different numerical approaches and codes for FDM exist. The most accurate of these are full wavefunction solvers  which solve the Schr\"odinger-Poisson equations for the complex wavefunction e.g. \citep{Schive2014, Mocz2019_PRL, MaySpringel2021, May.Springel_2023_HaloMass}.  \citet{MaySpringel2021} examines the impact of changing the dynamics between FDM and $N$-body CDM simulations (on fixed CDM initial conditions) in 10 Mpc/$h$ box at $z=3$ with $m=0.35$--$0.7 \times 10^{-22} \ \mathrm{eV}/c^2$.  \citet{May.Springel_2023_HaloMass} extends this analysis to also consider the impact of changing the initial conditions. They find that the initial conditions mostly impact the early evolution of the power spectrum and by $z=5$--$3$, non-linear growth has mostly made up for the initial suppression, with the non-linear FDM power spectrum gaining a positive bump on small scales dues to wave interference effects. \citet{Mocz2019_PRL} simulates the full dark matter wavefunction together with magneto-hydrodynamic effects in a 1.7 Mpc/$h$ box down to $z=5.5$ to study the impact on star formation.

\texttt{PyUltraLight} \citep{Edwards.etal_2018_PyUltraLightPseudospectral} and \texttt{UltraDark.jl} \citep{Glennon.etal_2023_SimulationsMultifield} are also pseudospectral wave solvers. \citet{Guo2021JCAP}, \citet{Glennon.etal_2023_SimulationsMultifield} and \citet{Luu2024MNRAS} also describe modelling multiple axion fields, motivated by the fact that the axion-like particles arising from the string axiverse \citep{Svrcek2006JHEP, Arvanitaki2010PhRvD} generically span a spectrum of masses. 

Instead of solving the Schr\"odinger-Poisson equations for the complex wavefunction, a variety of approaches leverage the fact that the Schr\"odinger-Poisson equations can be rewritten in the form of hydrodynamical equations and make use of Smoothed Particle Hydrodynamic (SPH) techniques to solve for the fluid variables. Using SPH in this way is advocated for by e.g. \citep{Mocz_2015_SP_SPH, Marsh_2015_NonlinearHydrodynamics}. One such code is \textsc{Ax-Gadget}  \citep{Nori_2018_AX-GADGET}, a modification of \textsc{Gadget-2} \citep{Springel_2005_CosmologicalSimulation} to accommodate for quantum pressure. SPH techniques can reach box lengths of $\sim \!\! 50 \ h^{-1}$ Mpc \citep{Zhang.etal_2018_ImportanceQuantum}, but are known to fail getting the details of the  interference patterns correct, as the fluid variables become ill behaved in regions of low density \citep{Mocz_2015_SP_SPH,Veltmaat2016PhRvD,  Hopkins_2019_numerics_FDM, SchwabeNiemeyer2022}. Different SPH approaches also disagree on the role of the quantum potential   with \citet{Nori_2018_AX-GADGET} finding that the quantum potential suppresses the matter power spectrum while \citet{Veltmaat2016PhRvD} finds enhancement on small scales.

Other techniques use a hybrid approach, such as \textsc{Axionyx} \citep{Schwabe.etal_2020_SimulatingMixed} that extends the \textsc{Nyx} code \citep{Almgren.etal_2013_NyxMassively} to solve systems with mixed fuzzy and cold dark matter, solving the FDM component via pseudospectral methods and running $N$-body for the CDM component. This was recently used in \citet{Lague.etal_2023_CosmologicalSimulations} for example, which simulates mixed CDM and FDM in 1--30 Mpc/$h$ boxes, with FDM masses of $10^{-25}$--$10^{-21} \ \mathrm{eV}/c^2$. \cite{SchwabeNiemeyer2022} also take this approach, solving $N$-body dynamics on large scales and full Schr\"odinger-Poisson on smaller scales to study individual halos. Both of these methods rely on translating the $N$-body particles into wavepackets at some point, as discussed in  \cite{Gough.Uhlemann_2022_MakingDark}. The Gaussian beam method was developed with the intention of reconstructing the wavefunction within a collapsed halo. To do this, $N$-body particles are regarded as phase-space wave-packets and augmented by a  complex phase that is co-evolved and coherent between neighbouring wave-packets. As the augmented particles move on classical trajectories, it is similar in spirit to the stationary phase approximation of the propagator \cite{Gough.Uhlemann_2022_MakingDark} which assigns classical wavefunctions to individual fluid streams without particle sampling. Both will produce interference effects in classically shell-crossed regions although the details of either method are not guaranteed to be correct compared to a full propagation of the wave-function. Neglecting all wave interference effects, \cite{Dome.etal_2023_CosmicWebElongation} introduced the term ``classical'' fuzzy dark matter, where the impact of the FDM is only captured by evolving suppressed initial conditions with standard $N$-body simulations. This was done in a 40 $h^{-1}$ Mpc box between redshifts $z\simeq 3.5$--$5.5$. As a result the final density fields cannot have interference patterns on any scale as seen in the lower left panel of Figure \ref{fig:slices-of-log-delta}.\\[0.1cm]

\subsection{Summary of theoretical approaches}
The computational expense of solving the Schr\"odinger-Poisson equations on cosmological scales has motivated a number of methods to make analytic or perturbative progress on such wavelike systems.

\citet{Li.etal_2019_NumericalPerturbative} attempts a perturbative expansion directly in the wavefunction $\delta\psi=\psi-\psi_{\rm bkgd}$, however they demonstrate that the requirement that the smallness of $\delta\psi$ generally breaks down before the smallness of fluid variables $\delta$, $\bm{u}$, limiting the applicability of this direct approach.

There are also approaches which are based around modifying  modelling tools used in the usual CDM context. \cite{Lague_2021_FDM_LPT} adjusts the particle displacements from Lagrangian Perturbation Theory (LPT) to account for the fact that linear growth in FDM is scale dependent. This is then applied to models of mixed dark matter. While this incorporates the quantum pressure at the linear level, it neglects non-linear effects in the displacement, as described in Appendix B of \cite{Uhlemann2014}, and fails to produce interference fringes in the final density field.

Approaches based on the Effective Field Theory of Large Scale Structure (EFTofLSS) have also been tried, such as in \cite{ManouchehriKousha.etal_2024_EffectiveField}. These approaches bundle the effects of the wave interference into the counterterms introduced in EFT (such as effective sound speed). This approach however relies on expanding the quantum potential as a series in powers of $\delta$, which breaks down where $\delta=-1$, see the quantum potential term defined in  equation \eqref{eqn:SP-fluid-eqns:Bernoulli}. Their simulations also follow the ``classical FDM'' approach of running $N$-body simulations on FDM suppressed initial conditions, failing to produce interference effects.

The approach used in this work is Propagator Perturbation Theory (PPT), which perturbatively solves for the propagator which describes how final wavefunctions are constructed from initial wavefunctions in terms of an effective potential. To lowest order, this approach describes free wave propagation in $a$-time, and at next-to-leading order the effective potential is time-independent, making these equations tractable to solve.

\section{Wavelike dark matter}

In this Section we sketch the relevant cosmological wave equations in this investigation, the true Schr\"odinger-Poisson equations for an ultralight non-relativistic scalar field and the Schr\"odinger equations which come out of the Propagator Perturbation formalism.

The de Broglie wavelength of associated to a particle of mass $m$ is given by
\begin{align}
\lambda_{\rm dB} &=  1.21 \ \mathrm{kpc} \left[\frac{10^{-22}\ \mathrm{eV}/c^2}{m}\right] \left[\frac{1 \rm \ km \ s^{-1}}{v}\right],
\end{align}
which motivates the definition $m_{22} = m/(10^{-22} \mathrm{eV}/c^2)$ as this sets the de Broglie wavelength on astrophysically relevant scales. Throughout this work we quote masses as values of $m_{22}$.

\subsection{Cosmological Schr\"odinger-Poisson equations}

The Schr\"odinger-Poisson equations \eqref{eqn:SP-full-eqns} can be recast into fluid-like variables via the Madelung representation \citep{Madelung1927}, allowing easier comparison to standard CDM dynamics. By writing $\psi = \sqrt{1+\delta}\exp(i m \phi /\hbar)$, the Schr\"odinger-Poisson equations become the continuity equation and a modified Bernoulli equation \citep{Chavanis_2012}:
\begin{subequations}\label{eqn:SP-fluid-eqns}
    \begin{align}
        &\del_t \delta + \frac{1}{a^2} \div{\left[(1+\delta) \grad \phi \right]} = 0 \,, \label{eqn:SP-fluid-eqns:continuity} \\
        &\del_t \phi + \frac{1}{2a^2} \abs{\grad \phi}^2 = - V_N  + \frac{\hbar^2}{2a^2}\frac{\nabla^2 \sqrt{1+\delta}}{\sqrt{1+\delta}}\,, \label{eqn:SP-fluid-eqns:Bernoulli} \\
        & \nabla^2 V_N = \frac{3\Omega_m^0 H_0^2}{2a}\delta\,, \label{eqn:SP-fluid-equns:Poisson}
    \end{align}
\end{subequations}
where $\delta$ is the density contrast and $\phi$ is the velocity potential which generates the peculiar velocity\footnote{Note that this is not the typical peculiar velocity $\bm{U}=\dv*{\xx}{\tau} = a\dv*{\xx}{t}$.} $\bm{u} = \dv*{\xx}{t} = \grad\phi$. The final term appearing in the Bernoulli equation \eqref{eqn:SP-fluid-eqns:Bernoulli} is a source of velocity dispersion which doesn't appear in the equations for a standard fluid, referred to as the ``quantum pressure'' or ``quantum potential''.

Linearising these equations in Fourier space leads to a second order equation for $\delta$ \citep{Chavanis_2012,Ferreira2021A&ARv}
\begin{equation}\label{eq:FDM-linear-growth}
\ddot{\delta}_k + 2 H(t) \dot{\delta}_k + \left[ \left(\frac{\hbar}{m}\frac{k^2}{2a^2}\right)^2 - \frac{3}{2}\Omega_{\rm FDM}(t)H^2(t)\right]\delta_k =  0,
\end{equation}
which sets the Jean's scale as \citep{Chavanis_2012,Marsh_2016_AxionCosmology}
\begin{equation}
k_J = 66.5 a^{1/4} \left(\frac{\Omega_{\rm FDM}^0h^2}{0.12}\right)^{1/4} m_{22}^{1/2} \ \rm Mpc^{-1}\,,
\end{equation}
where perturbations with $k<k_J$ grow in the linear regime while those with $k>k_J$ oscillate due to the quantum potential balancing against gravity. Solutions to equation \eqref{eq:FDM-linear-growth} determine the linear growth factors $D_\pm(k,a)$ for FDM which now no longer factorise into a spatial part and a time dependent part as they did with CDM. The exact growing mode solution to \eqref{eq:FDM-linear-growth} is \citep{Marsh_2016_AxionCosmology}
\begin{equation}
D_+^{\rm FDM}(k,a) = \frac{3\sqrt{a}}{\tilde{k}^2} \sin(\frac{\tilde{k}^2}{\sqrt{a}}) + \left[\frac{3a}{\tilde{k}^4} - 1\right]\cos(\frac{\tilde{k}^2}{\sqrt{a}} ) \,,
\end{equation}
where $\tilde{k}=k/\sqrt{m H_0}\propto k/k_J$. In the limit $\tilde{k}\ll 1$ this recovers $D_+^{\rm FDM}\sim a$ which is the usual CDM growing mode.

\subsection{Initial conditions}

Adding an additional scalar field to the content of the universe changes the initial conditions which set the linear power spectrum compared to a standard $\Lambda$CDM cosmology. The equations which couple the dynamics of this scalar field to the standard fluid content of the universe (CDM, baryons, neutrinos, photons etc) can be found in e.g. \citet{Hlozek.etal_2015_SearchUltralight}. For the purposes of structure formation, the principle effect of replacing the cold dark matter with an ultralight field is in the linear matter power spectrum, which is suppressed on small scales. This suppression is usually quantified relative to the standard CDM power spectrum such that
\begin{equation}
    P_{\rm FDM}(k,z) = T^2_{\rm FDM}(k,z) P_{\rm CDM}(k,z).
\end{equation}
The transfer function $T_{\rm FDM}(k,z)$ can be approximated by a redshift independent expression \citep{Hu.etal_2000_ColdFuzzy}
\begin{align}
    T_{\rm FDM}(k)  \approx \frac{\cos (x_J^3(k))}{1+x_J^8(k)}, \\ x_J(k) = 1.61 m_{22}^{1/18} k / k_{\rm J, eq}\,, \nonumber
\end{align}
where $k_{\rm J, eq} = 9 m_{22}^{1/2} \rm \ Mpc^{-1}$ is the Jeans scale at matter-radiation equality. Alternatively, the power spectrum of the axion field can be calculated by directly solving the Boltzmann hierarchy with the axion fluid equations included, as in the code \texttt{axionCAMB}\footnote{\url{https://github.com/dgrin1/axionCAMB}}\citep{Hlozek.etal_2015_SearchUltralight}.

\subsection{Propagator Perturbation Theory}\label{sec:PPT-theory}

Propagator Perturbation Theory introduces a semiclassical wavefunction $\psi$ to describe the dark matter field, using a semiclassical parameter, $\hbar_{\rm PPT}$, which controls the amount of wave behaviour in the system, akin to the combination $\hbar/m$ in the FDM scenario. In the semiclassical limit $\hbar_{\rm PPT}\to 0$ the dynamics of 1PPT reproduce Zel'dovich approximation (1LPT) dynamics, and can straightforwardly extract quantities in Eulerian space. At higher order, 2PPT produces similar dynamics to 2LPT in the semiclassical limit, but can also protect certain quantities such as the generation of spurious vorticity to higher perturbative order \citep{Uhlemann2019}. PPT was established in \cite{Uhlemann2019}, extended to two cold fluids in \cite{Rampf2021MNRAS, Hahn_2021_ICs_2fluid} and applied as an in the context of generating initial conditions for cosmological simulations in \cite{Michaux.etal_2021_AccurateInitial}. 

While this formalism is fundamentally based on different underlying physical assumptions than ``true'' Schr\"odinger-Poisson/FDM systems discussed above, it does retain non-linear wavelike phenomena in its evolution, rather than relying on classical fluid evolution or simply changing linear growth. Wave interference has certain universal features which even approximate schemes such as PPT can capture \citep{Gough.Uhlemann_2022_MakingDark}. Density fields created out of wavefunctions rather than tracer particles are also particularly useful for extracting information from underdense environments such as the Lyman-$\alpha$ forest \citep{Porqueres_2020}, due to the wavefunction's uniform resolution in position space.

The PPT equations are usually written in units such that $4\pi G\bar{\rho}_0 = \frac{3}{2}$ (equivalent to setting $\Omega_m^0 H_0^2 = 1$). Here we keep the physical constants more explicit to aid in mapping between PPT results and more standard PT results. 

The dynamical equations of PPT are
\begin{subequations}\label{eqn:PPT-schrodinger-poisson}
\begin{align}
    i\partial_a \hbar_{\rm PPT}\psi &= -\frac{\hbar_{\rm PPT}^2 }{2}\nabla^2 \psi + V_{\rm eff} \psi\,, \label{eqn:PPT-schrodinger} \\
    V_{\rm eff} &= \frac{3 \Omega_m^0 H_0^2}{2a}(\phi_v + \varphi_g)\,, \\
    \nabla^2 \varphi_g &= \frac{1}{a}\left(\abs{\psi}^2-1\right)\,, \label{eqn:PPT_poisson}
\end{align}
\end{subequations}
where the effective potential encodes both the gravitational effects and expansion effects through the Hubble drag in the velocity potential $\phi_v$.\footnote{Note that we take the sign convention of \cite{Gough.Uhlemann_2022_MakingDark} where $\bm{v} = \grad\phi_v$ rather than the convention of \cite{Uhlemann2019} where $\bm{v} = -\grad\phi_v$.}  It is this effective potential $V_{\rm eff}$ which is solved for perturbatively in PPT. Notice that PPT takes the scale factor $a$ to be its natural time variable, rather than cosmic-time $t$\footnote{See \citet{Rampf2021MNRAS} for generalisation to $D$-time where $D$ is the linear growth factor.} (see Appendix \ref{app:sec:matching_PPT_to_LPT} and Appendix B of \cite{Uhlemann2019} for further discussion of the time variables).

Under the Madelung transformation $\psi = \sqrt{1+\delta}\exp(i\phi_v/\hbar_{\rm PPT})$, these equations become the following fluid equations
\begin{subequations} \label{eqn:PPT-fluid-equations}
    \begin{align}
        &\partial_a \delta + \grad \cdot [(1+\delta)\grad\phi_v] = 0\,,\label{eqn:PPT-continuity}\\
        &\partial_a \phi_v + \frac{1}{2} \abs{\grad\phi_v}^2 = -V_{\rm eff} + \frac{\hbar_{\rm PPT}^2}{2} \frac{\nabla^2 \sqrt{1+\delta}}{\sqrt{1+\delta}}\,,  \label{eqn:PPT-Bernoulli}\\
        &\nabla^2 \varphi_g = \frac{\delta}{a}\,. \label{eqn:PPT-poisson-fluid}
    \end{align}
\end{subequations}

The initial conditions for PPT (or LPT) are determined by requiring analytic behaviour at $a=0$. By considering the fluid equations as $a\to 0$, the initial conditions for the density contrast and velocity potential are tethered to each other \citep{Brenier2003} 
\begin{equation}\label{eqn:ZA_boundary_conditions}
    \delta^{\rm (ini)} = 0\,, \quad \phi_v^{\rm (ini)} = - \varphi_g^{\rm (ini)}\,.
\end{equation}
These boundary conditions select for the growing-mode solutions and are vorticity free. In these coordinates, the effective potential becomes very simple. To lowest perturbative order, the effective potential vanishes, making the equations of 1PPT simply a free Schr\"odinger equation (in $a$-time). At NLO, the effective potential is non-zero, but is ($a$)-time independent, making these dynamical equations straightforward to solve.

The free Schr\"odinger/1PPT dynamics where this effective potential vanishes asymptotically describe the same dynamics as 1st order Lagrangian Perturbation Theory (LPT) or the Zel'dovich approximation \citep{Zeldovich1970}, where particles are given velocities (relative to $a$-time) set by the initial gravitational potential, then free stream until shell crossing. This has also been studied as the ``free-particle Schr\"odinger equation'' \citep{ColesSpencer2003, ShortColes2006, Gallagher_2022_SPvoids}. The correspondence of the free wavefunction of 1PPT to  Zel'dovich approximation dynamics is explicitly demonstrated in \cite{Gough.Uhlemann_2022_MakingDark}. Note that within both PPT and LPT vorticity and velocity dispersion will be generated through shell crossing and the formation of multi-streaming regions. In LPT recovering those Eulerian quantities requires an explicit averaging over distinct fluid streams. PPT instead provides direct access to them through the phase information in the wave function, see e.g. Figure~5 in \cite{Uhlemann2019} and Figures~10~and~14 in \cite{Gough.Uhlemann_2022_MakingDark}. Note that the absence of secondary infall in a single time-step evolution according to LPT and PPT limits the number of attainable fluid streams.

\section{Numerical methods for perturbative forward modelling}\label{sec:howw-fuzzy-numerics}

To examine the effects of initial conditions and evolution dynamics we produce a set of perturbative forward models using a mildly modified version of the initial conditions code \texttt{monofonIC}\footnote{\url{https://bitbucket.org/ohahn/monofonic/src/master/}} \citep{Hahn_2021_ICs_2fluid, Michaux.etal_2021_AccurateInitial}. This is intended to generate initial conditions for cosmological simulations using either $n$LPT to generate particle displacements or $n$PPT to generate the initial density and velocity fields. For our analysis we run \texttt{monofonIC} in 1-fluid mode (dark matter only) with the cosmological parameters shown in Table \ref{tab:cosmo_params_for_monofonic} in a box of $(128 \ h^{-1} \ \rm Mpc)^3$ on a $(1024)^3$ grid. For the LPT runs we run with one particle per grid point.  Note that while baryon, radiation, and neutrino fractions are included in Table \ref{tab:cosmo_params_for_monofonic}, they are used only in setting the appropriate linear power spectrum, the final output field is dark matter only.

\begin{table}[h!t]
    \centering
    \begin{tabular}{ccccc}
        \hline
         $\Omega_m$  &   $\Omega_b$  & $\Omega_c$ & $\Omega_r$ & $h$\\
         0.3158 & 0.0494 & 0.264979 & $7.99185\times 10^{-5}$ & 0.67321 \\

        \hline
        \hline
         $\sum m_\nu$ [$\mathrm{eV}/c^2$] & $\Omega_\nu$ &  $n_s$ & $A_s$ & $\sigma_8$\\
         0.06 & 0.001423 & 0.9661 & $2.094\times 10^{-9}$ & 0.8102\\
        \hline
    \end{tabular}
    \caption[Cosmological parameters used to produce density fields from \texttt{monofonIC}.]{Cosmological parameters used to produce our density fields with \texttt{monofonIC}. We set the value of  $A_s$ rather than fixing $\sigma_8$, the value of $\sigma_8$ is the extrapolated value at $z=0$.}
    \label{tab:cosmo_params_for_monofonic}
\end{table}

The output density fields from \texttt{monofonIC} are generated in the following steps:
\begin{enumerate}
	\item Provide a transfer function for the total matter at the target output redshift $T_m(k,z_{\rm t})$ to \texttt{monofonIC}. 
	\item Generate a Gaussian random field with a power spectrum matching the provided target transfer function, call this $\delta_{\rm code}(\xx,a_{\rm t})$. As we're in the perturbative limit, this factorises into $\delta(\xx,a) = D_+(a)C_+(\xx)$. The spatial part of this product is solved by
	\begin{equation}
	C_+(\xx) = \frac{\delta_{\rm code}(\xx,a_{\rm t}) }{D_+(a_{\rm t})}\,,
	\end{equation}
	\item The initial gravitational potential is found by solving the Poisson equation for $\varphi_g$ then backscaling by the CDM linear growth factor taking the $a\to 0$ limit
	\begin{equation}
	\varphi_g^{\rm (ini)}(\xx) = \frac{\nabla^{-2}\delta_{\rm code}}{D_+(a_{\rm t})} \lim_{a\to 0}\frac{D_+(a)}{a}.
	\end{equation}
	\item This initial gravitational potential is then used in the relevant PT scheme (either LPT or PPT) to determine the displacement field or the initial wavefunction and effective potential. The perturbative evolutions are then evolved from these initial conditions to the ``start redshift'' (which will be the final output of this process). In the PPT case the density is constructed as simply $\abs{\psi}^2$ on this grid. 
	\item For the classical dynamics (LPT) runs, the density fields are constructed from particle displacements via \texttt{Pylians3}\footnote{\url{https://pylians3.readthedocs.io/en/master/index.html}} \citep{Pylians} using the cloud-in-cell mass assignment scheme. 
	
\end{enumerate}

Slices of the final density fields at $z=4$ are shown in Figure~\ref{fig:slices-of-log-delta}. All our output fields have the same fixed target redshift of $z=4$ and varying ``start redshifts''\footnote{The term ``start redshift'' is the name given by \texttt{monofonIC}, as its standard use is in setting initial conditions for $N$-body simulations, which would start at $z_{\rm start}$. When used as a forward model, this ``start redshift'' is the output redshift of interest, as all the perturbative schemes ``start'' at $a=0$.} to investigate the time evolution of certain quantities. All runs which we compare directly are run from the same random seed, providing the same white noise field which is scaled by the target transfer function.

The CDM and FDM linear power spectra which are supplied to \texttt{monofonIC} were calculated using  \texttt{axionCAMB}\footnote{\url{https://github.com/dgrin1/axionCAMB}} \citep{Hlozek.etal_2015_SearchUltralight}, an extension to the standard Boltzmann solver \textsc{CAMB} \citep{Lewis.Bridle_2002_CosmologicalParameters}. As \texttt{axionCAMB} requires both axions and standard cold dark matter, the fully CDM/FDM transfer functions were calculated by using an axion fraction of $10^{-7}$ and $(1-10^{-7})$ respectively. These will be referred to as the CDM ICs and FDM ICs throughout. The FDM power spectra were specified with a boson mass of $m_{22}=0.1$ to accentuate the wave effects on the final field while facilitating a comparison to existing large-scale full FDM simulations \citep{MaySpringel2021,May.Springel_2023_HaloMass}. 

By default, if \texttt{monofonIC} calculates PPT fields, it will choose the minimum value of $\hbar_{\rm PPT}$ that is allowed by the initial gravitational potential to avoid aliasing, in order to match as closely to CDM results as possible. We have modified the \texttt{monofonIC} code to allow specification of the size of $\hbar$ which can be larger than this. As \texttt{monofonIC} specifies the box size in $h^{-1} \ \rm Mpc$ units, the value of $\hbar_{\rm PPT}$ is always specified in $(h^{-1} \ {\rm Mpc})^2$. We choose to run on a fixed size of $\hbar_{\rm PPT}$ which is about 10\% larger than the minimum value in the CDM ICs, which ensures that the same $\hbar_{\rm PPT}$ on different random seeds or different initial conditions would also avoid aliasing.\footnote{The minimum value of $\hbar_{\rm PPT}$ on the random seed used in this paper was 1.04619 for CDM ICs and 0.987302 on FDM ICs.}

We choose $\hbar_{\rm PPT} = 1.1 \ (\mathrm{Mpc}/h)^2$, which has a similar effect on the power spectrum as a boson mass of $m_{22}=0.1$ on perturbative scales, as shown in Figure~\ref{fig:Pk-ratios-with-data}. As generally $\hbar_{\rm PPT}$ is inversely related to the FDM particle mass, this is the closest $\hbar_{\rm PPT}$ which was feasible to give effects similar to reasonably physical FDM masses. Both parameters $\hbar_{\rm PPT}$ and $\hbar/m_{22}$ control the size of a quantum potential term that is sourced by the curvature of density fluctuations, see equations~\eqref{eqn:SP-fluid-eqns:Bernoulli}~and~\eqref{eqn:PPT-Bernoulli}. As they originate from different time and velocity variables, they have a different time-dependence 
$\hbar_{\rm PPT}^2\propto (\hbar/ m_{22})^{2}  a^{-3}$ such that matching their integrated impact depends on the target redshift and the onset of relevance of the quantum potential. We discuss the matching between PPT and FDM in more detail in Appendix \ref{app:sec:matching_PPT_to_LPT}, and present the effect of a larger $\hbar_{\rm PPT}$ in Appendix \ref{app:sec:largerhbar}.

Our choice of a single FDM species with $m_{22}=0.1$ should be regarded as a toy model that is compatible with constraints from CMB+galaxy clustering \citep{Hlozek.etal_2015_SearchUltralight, Hlozek2018MNRAS, Rogers2023JCAP}, but incompatible with observations from  Lyman-alpha \citep[][limiting $m_{22}\gtrsim10$]{Irsic2017PhRvL,Armengaud2017MNRAS, Kobayashi2017PhRvD, PhysRevLett.128.171301}, galaxy UV luminosities \citep[][constraining $m_{22}\gtrsim 0.4$]{Winch2024arXiv} and ultrafaint dwarf galaxies \citep[][constraining $m_{22}\gtrsim 10^3$]{Dalal2022}. Going to higher masses $m_{22}$ requires smaller $\hbar_{\rm PPT}$, and thus a finer grid at fixed box size thus increasing memory requirements while the perturbative evolution remains computationally cheap. 
These mass constraints can be relaxed or evaded if FDM makes up only a fraction of the dark matter content \citep{Gosenca2023,Lague.etal_2023_CosmologicalSimulations,Huang2023} or in the presence of self-interaction terms \citep{Desjacques2018,Leong2019,Mocz2023,Painter2024}. In the context of growing modes in perturbation theory, two dark matter species coupled through the gravitational potential can be realised by imprinting the initial relative densities on to the particle masses (LPT) and the wave function amplitude (PPT), and evolving them with the displacements (LPT) and propagator (PPT) for a single species
\citep[in analogy to a two-fluid system with dark matter and baryons discussed in detail in][]{Rampf2021MNRAS,Hahn_2021_ICs_2fluid}. This could be applied to evolve a mixed CDM and FDM model using a combination of LPT and PPT. Realising mixed FDM models could be achieved by generalising existing results for two-species PPT with a common $\hbar_{\rm PPT}$ to differing $\hbar_{\rm PPT}^\alpha$ for individual species $\alpha$.

\section{Statistics of the density field}

In this Section we present statistics measured on the density field produced from \texttt{monofonIC} as described above. In principle one could also calculate statistics on the velocity fields predicted with these forward models (directly in LPT and through the wavefunction phase in PPT), which would be  a natural extension to this work.

As much as possible we stick to the same colouring scheme for the different statistics investigated in this Section. Colour (or marker shape) is used to differentiate perturbation scheme, representing whether wave dynamics are present or not. The line style is used to differentiate initial conditions, with solid lines for CDM ICs and dashed for FDM ICs. Transparency is used in a few plots  to indicate perturbation theory order.

\subsection{Power spectrum}
\begin{figure}[t]
    \centering
\includegraphics[width=\columnwidth]{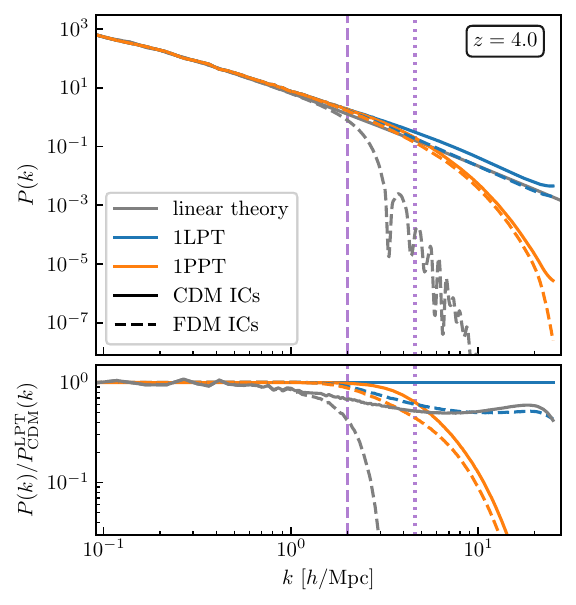}
    \caption[Four way comparison of the matter power spectrum.]{Four way comparison of the power spectrum from 1LPT (CDM) and 1PPT (wave) perturbative simulations with either FDM or CDM initial conditions in a $L=128 \ \mathrm{Mpc}/h$ box. The target linear power spectrum for CDM/FDM is shown as the grey lines. The lower plot shows the ratio of the power spectrum to the classically evolved CDM initial conditions power spectrum. This plot uses the same colouring as Figure~6 in \cite{May.Springel_2023_HaloMass} for easy comparison. The dashed vertical line marks the $k$ value where 2LPT produces the most non-linear power compared to 1LPT, and the dotted vertical line denotes the $k$ value beyond which  2LPT produces less power than 1LPT. These provide the bounds on the region of validity for perturbation theory.}
    \label{fig:Pk-z4}
\end{figure}
\begin{figure}
    \centering
\includegraphics[width=\columnwidth]{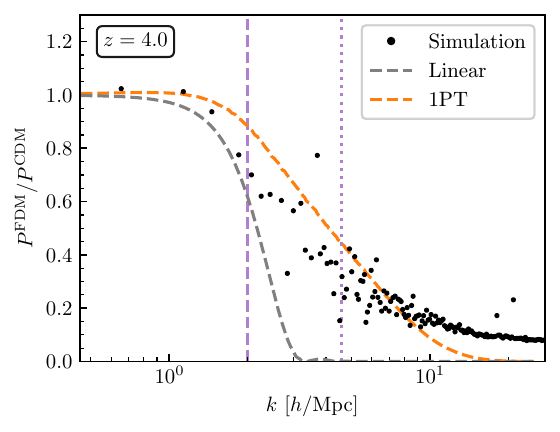}
    \caption[]{Ratios of the wavelike to classical power spectrum for three cases. First, the ratio of a full Schr\"odinger-Poisson simulation with suppressed initial conditions ($m_{22}=0.1$) to an $N$-body simulation with cold initial conditions \citep[black points,][]{May.Springel_2023_HaloMass}. Second,, power spectrum from 1PPT + FDM ICs ($\hbar_{\rm PPT}=1.1$ (Mpc$/h$)$^2$, $m_{22}=0.1$) to 1LPT + CDM ICs (orange dashed). The vertical lines show the relevant perturbative scales as in Figure \ref{fig:Pk-z4}. Third, the linear FDM power spectrum to the linear CDM power spectrum for $m_{22}=0.1$ (dashed grey).}
    \label{fig:Pk-ratios-with-data}
\end{figure}

The the dependence of the power spectrum on initial conditions and dynamics in a full Schr\"odinger-Poisson solver has been presented in e.g. Figure 6 of \cite{May.Springel_2023_HaloMass}. In Figure~\ref{fig:Pk-z4} we plot the measured power spectrum at $z=4$ in a four way comparison between the 1LPT/1PPT dynamics and the CDM/FDM initial conditions, as well as the linear theory power spectra produced by \texttt{axionCAMB}. At $z=4$ the effect of the initial conditions has largely been removed, as non-linear growth has enhanced the initially suppressed power from the FDM initial conditions. Relative to the fully classical evolution (1LPT + CDM ICs, blue solid line) propagation with wave dynamics (orange lines) still features a suppression of power on large scales, owing to the quantum potential term in equation \eqref{eqn:PPT-Bernoulli}. The vertical lines in Figure \ref{fig:Pk-z4} indicate a rough region of validity for our perturbation theory. The dashed vertical line marks the scale where 2LPT provides the most non-linear power compared to 1LPT, while the dotted vertical line marks the scale where 2LPT produces less non-linear power compared to 1LPT.  We see less enhancement over linear power than in the full Schr\"odinger-Poisson simulations in \citet{May.Springel_2023_HaloMass}, however the ratio between the power in the fully wave-based and fully classical cases is similar on the scales of interest, as shown in Figure \ref{fig:Pk-ratios-with-data}.

The ratio between power spectra from fully wavelike and fully classical density fields is used to evaluate whether the chosen value of  $\hbar_{\rm PPT}=1.1 \ h^{-2} \ \rm Mpc^2$ is well matched to the mass used to set the FDM initial conditions, $m_{22}=0.1$ in Figure \ref{fig:Pk-ratios-with-data}. The simulation points are from a full Schr\"odiner-Poisson simulation with FDM initial conditions (with $m_{22}=0.1$) in comparison to an $N$-body simulation with CDM initial conditions from \cite{May.Springel_2023_HaloMass}. These full simulations were performed in a 10 Mpc/$h$ side length box with $(4320)^3$ grid, the $N$-body simulation using $(2048)^3$ particles. The relative suppression from perturbation theory with this choice of $\hbar_{\rm PPT}$ slightly underestimates the suppression effect due to wave interference compared to the non-linear simulations. This indicates our choice of $\hbar_{\rm PPT} = 1.1 \ h^{-2} \ \rm Mpc^2$ represents a conservative estimate of the impact of wave dynamics for a FDM particle mass of $10^{-23} \ \mathrm{eV}/c^2$. Since the precise mapping between $\hbar_{\rm PPT}$ and the FDM particle mass requires further theoretical understanding (see Appendix \ref{app:mappinghbar} for a discussion and Appendix \ref{app:sec:largerhbar} for clustering statistics at larger $\hbar_{\rm PPT}$) we proceed with this choice for the remainder of the analysis, acknowledging that this may underestimate the impact of interference on the density field.

\subsection{Matter PDFs and skewness}

The distribution of matter density in spheres of radius $R$ is a particularly simple non-Gaussan statistic. We use spherical top-hat filters in real space to smooth the density field, corresponding to the Fourier space window function
\begin{align}
    \delta_R(k) &= \tilde{W}(kR) \delta(k)\,, \\
    \tilde{W}(kR) &= \frac{3}{(kR)^3}\left(\sin(kR)-kR\cos(kR)\right). \nonumber
\end{align}
The matter PDF in with top-hat smoothing is well modelled in the quasilinear regime by large deviations theory \citep{Bernardeau14,Uhlemann16} valid for a wide range of cosmologies even beyond $\Lambda$CDM \citep{Uhlemann:2020,Cataneo.etal_2022_MatterDensity}.

\begin{figure}[h!t]
    \centering
    \includegraphics[width=\columnwidth]{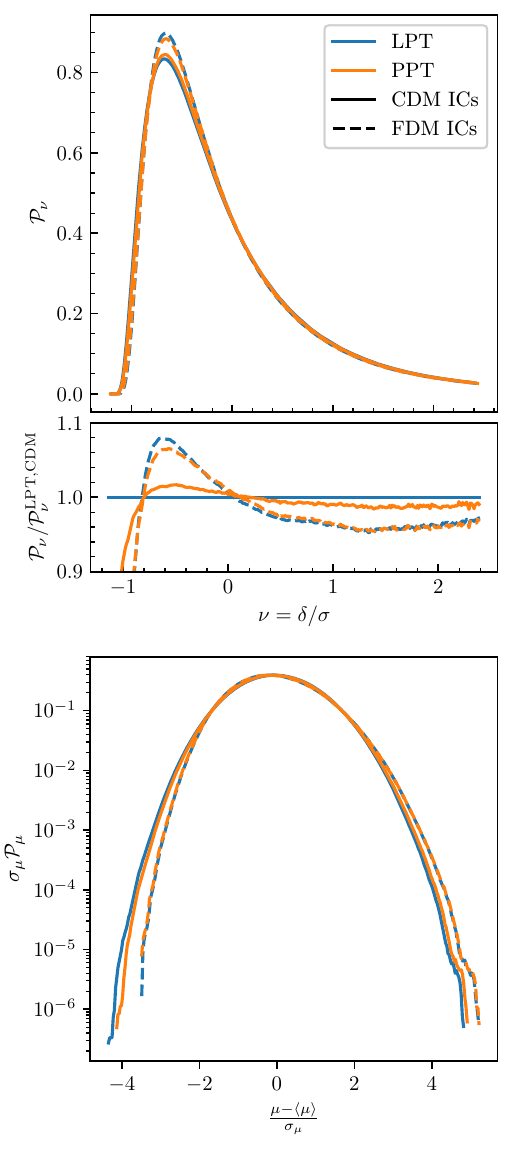}
    \caption[Four way comparison for the matter PDF in spheres of $R= 1 \ h^{-1} \ \rm Mpc$ at $z=4$.]{Four way comparison for the matter PDF in spheres of $R= 1 \ h^{-1} \ \rm Mpc$ at $z=4$.  (Top panel) The PDF of the rarity $\nu=\delta_R/\sigma(R)$ where $\sigma^2(R)$ is the variance of the density field on scale $R$. (Middle panel) The ratio of the rarity PDFs to the fully classical (LPT + CDM ICs) PDF.  (Bottom panel) The PDF of the rarity relative to the log density $\mu = \log(1+\delta)$. Note that suppressed initial conditions (dashed lines) introduce additional skewness to the PDF.  }
    \label{fig:matter_pdf_mu_nu_r1_z4}
\end{figure}

Figure \ref{fig:matter_pdf_mu_nu_r1_z4} shows the PDF of the rareness of the density $\nu=\delta/\sigma$ and the log-density $\mu=\log(1+\delta)$ while varying the evolution dynamics and the initial conditions in spheres of radius $R= 1 \ h^{-1} \ \rm Mpc$ (8 times the grid scale). As changing the initial conditions changes the final variance on scale $R$, by plotting the PDF of the rareness the PDFs are normalised to have the same width. After normalising the width of the PDF it becomes evident that the FDM IC simulations (dashed lines) have an additional skewness compared to the simulations run on CDM initial conditions at $R= 1 \ h^{-1} \ \rm Mpc$. Changing the dynamics on this scale doesn't have a large impact, with the main impact being produced by changing the initial conditions.  

The skewness seen in these PDFs can be well predicted by perturbation theory in terms of the variance. The \textit{reduced cumulants}
\begin{equation}
S_n = \frac{\ev{\delta^n}_c}{\ev{\delta^2}_c^{n-1}}\,,
\end{equation}
are particularly robust hierarchical ratios \citep{Bernardeau_1994_SkewnessKurtosis,Bernardeau:2002}. In the case of an unsmoothed density field and an EdS universe the reduced cumulants are constant to tree order in standard perturbation theory. When smoothed with a top-hat filter, the reduced cumulants acquire a correction which depends only on the linear variance. SPT predicts the reduced skewness $S_3(R)$ \citep{Bernardeau_1994_EffectsSmoothinga}
\begin{equation}
S_3^{\rm tree, SPT}(R) = \frac{34}{7} + \dv{\log\sigma_{\rm L}^2(R)}{\log R}\,.
\end{equation}
However, as the density fields considered in our analysis are not full non-linear simulations the recovered reduced skewness will differ from this value, owing to the difference in dynamics. It is known that $n$LPT only accurately reproduces cumulants up to $S_{n+1}$ \citep{Munshi.etal_1994_NonlinearApproximations}. The calculation for reduced skewness can be adapted to the Zel'dovich approximations/1LPT by replacing the 2nd order perturbation kernel $F_2^{\rm SPT} \to F_2^{\rm ZA}$ in the calculation of $\ev{\delta^3_R}$ where \citep{Scoccimarro.Frieman_1996_LoopCorrections} 
\begin{align}
F_2^{\rm SPT}(\kk_1, \kk_2) &= \frac{5}{7} + \frac{1}{2}\frac{\kk_1\cdot\kk_2}{k_1 k_2}\left(\frac{k_1}{k_2} + \frac{k_2}{k_1}\right) + \frac{2}{7}\frac{(\kk_1\cdot \kk_2)^2}{k_1^2 k_2^2}, \\
F_2^{\rm ZA}(\kk_1, \kk_2) &= \frac{1}{2}\frac{(\kk_1+\kk_2)\cdot(\kk_1+\kk_2)}{k_1^2 k_2^2}.
\end{align}
This replacement does not affect the smoothing term, only the bare value of the skewness, leading to \citep{Bernardeau1995ApJ}
\begin{equation}
S_3^{\rm tree, ZA}(R) = 4 + \dv{\log\sigma_{\rm L}^2(R)}{\log R}.
\end{equation}
This result can also be derived via vertex generating functions  \citep{Bernardeau1995ApJ}.
As the linear variance is an integral over the linear power spectrum
\begin{equation}
\sigma_{\rm L}^2(R) = \int \frac{\dd{k}}{2\pi^2} W(kR)^2 P_{\rm L}(k)\,,
\end{equation}
the variance for the field run on FDM initial conditions at a fixed physical scale $R$ will be smaller than the variance in the field run on CDM due to the suppression in linear power.

\begin{figure}[t]
    \centering\includegraphics[width=0.5\textwidth]{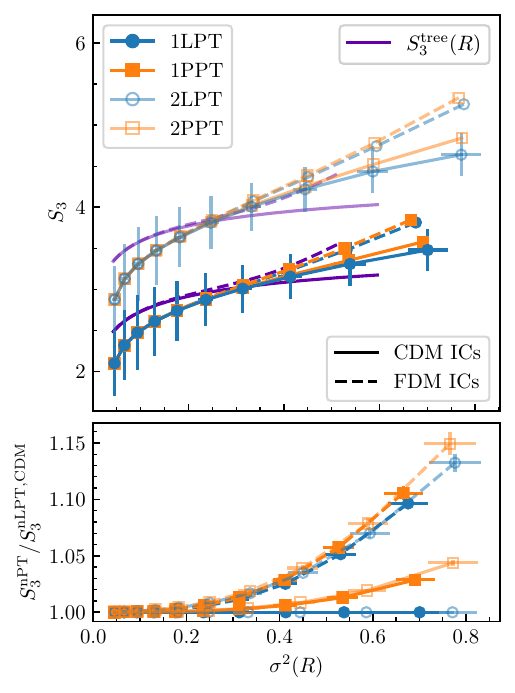}
    \caption[The reduced skewness measured with fluid and wave evolution on CDM and FDM initial conditions.]{(Top panel) Reduced skewness as a function of  variance at $z=4$, measured on top-hat smoothing scales $1$--$10 \ h^{-1} \ \rm Mpc$, log-spaced in $R$. Paler lines/open symbols show the results from 2nd order PT. The error bars are the standard deviation of the measured quantities across 8 subboxes. The errors are only shown on one line on the top plot to reduce clutter, but errors on other lines are similar. While the errors are relatively large (comparable with the scatter from measuring this across different realisations), the relative splitting (shown in the lower panels) have smaller scatter across realisations, demonstrating that wave effects generally enhance $S_3$ compared to LPT and CDM, in both smoothing ICs and dynamics. The purple lines (with no markers) are the tree order predictions for $S_3(R)$, which depend only on the linear variance. }
    \label{fig:S3_1column}
\end{figure}

Figure~\ref{fig:S3_1column} shows the measured reduced skewness against the measured variance $\sigma^2(R)$ on physical scales between 1--10 Mpc/$h$. The paler lines/open symbols show that the 2nd order PT simulations recover the vertical shift corresponding to $S_3^{\rm tree, SPT}=34/7$ instead of the $S_3^{\rm tree, ZA}=4$. We see also that on small scales (higher variance) the FDM initial conditions produce a higher reduced skewness, as anticipated from the PDFs in Figure~\ref{fig:matter_pdf_mu_nu_r1_z4}. While the standard deviation across subboxes of the simulation is (as indicated by the errorbars) is comparable with this enhancement, the ratio to the LPT+CDM ICs simulation is more robust, indicating that while the entire $S_3$ line has scatter comparable to the errorbars shown across realisations, the relative enhancement seen is not simply numerical noise. However we can see that the enhancement in $S_3$ due to the initial conditions appears to be in line with the amount of enhancement anticipated from the smoothing term $\dv*{\log \sigma_{\rm L}^2}{\log R}$. This means that the change in the skewness is largely captured by the change in the suppression from the change in the linear power spectrum. There is a small enhancement to the skewness by introducing wave dynamics but this is subdominant to the change in the overall variance from the initial conditions. This agrees with the findings of \cite{Dome.etal_2023_CosmicWebDissection} in their classical FDM simulations, where they find simulations with higher suppression in their initial conditions have higher values of $S_3$, even on scales significantly smaller than considered here.

The enhancement of non-Gaussian features such as the (reduced) skewness for fields with less small scale power is expected by considering the dynamics of the smoothed density field in Lagrangian space \citep{Bernardeau_1994_SkewnessKurtosis, Bernardeau1995ApJ}. At a fixed smoothing scale $R$, regions which are overdense in the final field evolved from larger regions via collapse, while underdense grow from smaller regions. Thus in fields with less small scale power, this asymmetry between over- and underdense regions is stronger than fields with power on small scales, enhancing the skewness as seen. This effect is similar to changing the effective tilt in the primordial power spectrum, as for a linear power spectrum $P_{\rm L}(k)\propto k^n$, the tree order skewness becomes
\begin{equation}
S_3^{\rm tree}(R) = S_3^{\rm tree}(0) + n(R).
\end{equation}
The removal of small scale power by smoothing on order Mpc scales is known to improve the accuracy of the Zel'dovich approximation by reducing the amount of shell crossing, known as the \emph{truncated Zel'dovich approximation} \citep{Melott1994_TruncatedZA}.

From the matter PDF and the reduced skewness it appears that the averaging of the density in cells efficiently erases most of the dynamical difference between LPT and PPT, at least in this perturbative regime. However, exactly how strongly separated the dynamical and initial condition effects on $S_3$ are in PPT depends on understanding the $\hbar_{\rm PPT}$ to mass mapping better, as a larger $\hbar_{\rm PPT}$ can cause enhancement similar to the FDM initial conditions even from cold initial conditions, generating the skewness purely through dynamics (see Appendix \ref{app:sec:largerhbar}).

\subsection{Critical points of the density field}

Elements of the cosmic web can be defined and identified in a variety of ways.  These elements are usually referred to by their visual nature, knots/peaks for the compact and densest points, line like filaments which extend in one dimension, and  sheets or walls which bound the underdense voids.  The NEXUS/NEXUS+ algorithm \citep{Cautun.etal_2013_NEXUSTracing} assigns a cosmic web element to every point in a volume a classification based on multiscale analysis of the Hessian of cosmic fields. This was applied in the context of classical FDM in \cite{Dome.etal_2023_CosmicWebDissection}. Classifications based on different fields, such as the tidal or velocity shear fields give rise to the cosmic T-web \citep{Forero-Romero.etal_2009_DynamicalClassification, Aycoberry.etal_2023_TheoreticalView} and V-web \citep{Hoffman.etal_2012_KinematicClassification, Cui.etal_2018_LargescaleEnvironment}. Other classifications include identifying persistent topological structures via Morse methods \citep{Colombi.etal_2000_TreeStructure, Sousbie_2011_PersistentCosmic, Sousbie.etal_2011_PersistentCosmic}, segmenting the density field  \citep[the SpineWeb formalism, ][]{Aragon-Calvo.etal_2010_SpineCosmic}, and counting phase space folds \citep[the ORIGAMI method, ][]{Falck.etal_2012_ORIGAMIDelineatinga}.

\begin{figure}[h!t]
\centering
\includegraphics[width=\columnwidth]{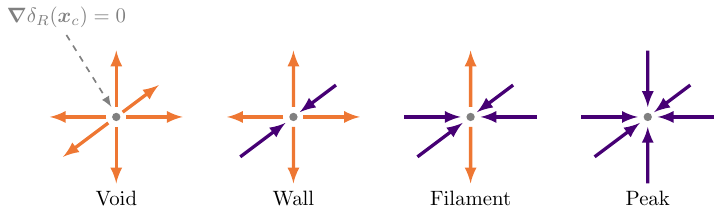}
\caption[Schematic diagram for critical point classification. ]{Classification of cosmic web elements via critical points of the smoothed density field. The directions/colours of the arrows indicate the sign of the eigenvalues $\lambda$ of the Hessian matrix $H_{ij}$ evaluated at the critical point. Outwards/orange arrows correspond to $\lambda>0$ while inwards/purple arrows correspond to $\lambda<0$.}
\label{fig:critical_points}
\end{figure}

For our purposes we concern ourselves with critical points of the smoothed density field, $\delta_R$.\footnote{In this Section we continue to smooth the density field with spherical top-hat filters, as in the previous section. This is in contrast to many other critical point studies which use a Gaussian smoothing filter.} The Hessian matrix of the smoothed density field
\begin{equation}
    H_{ij}(\xx) = \nabla_i\nabla_j \delta_R(\xx)\,,
\end{equation}
encodes the curvature of the density field in 3 dimensions. Critical points of the density field (defined by $\grad\delta_R(\xx_c) = 0$), can be classified by the number of positive eigenvalues of $\mathsf{H}$ at that point, leading to the classification (schematically represented in Figure \ref{fig:critical_points}):
\begin{itemize}
    \item 0 positive eigenvalues, Peak/Node/Knot
    \item 1 positive eigenvalues, Filament
    \item 2 positive eigenvalues, Wall
    \item 3 positive eigenvalues, Void
\end{itemize}
Note that in other critical point analyses  \citep[e.g. ][]{Forero-Romero.etal_2009_DynamicalClassification, Hoffman.etal_2012_KinematicClassification, Carlesi.etal_2014_HydrodynamicalSimulations, Cui.etal_2018_LargescaleEnvironment, Libeskind.etal_2018_TracingCosmic, Suarez-Perez.etal_2021_FourCosmic, Aycoberry.etal_2023_TheoreticalView} a small non-zero threshold for the eigenvalues of the relevant matrix is used for better visual agreement between the field and the located cosmic web elements. 

We run the extrema finding code \texttt{py\_extrema}\footnote{\url{https://github.com/cphyc/py_extrema}} (\citet{Shim.etal_2021_ClusteringCritical}, see also Appendix G of \citet{Gay.etal_2012_NonGaussianStatistics}) on the density fields produced by \texttt{monofonIC} after smoothing. This calculates the derivatives and Hessian of the smoothed density field in Fourier space and then discards multiple critical points of the same kind which are found to lie in the same pixel. As the derivatives are calculated in Fourier space and require periodic boundary conditions we have to compute them on the full field. The smoothed density field is downsampled from a $(1024)^3$ grid to $(256)^3$ grid before searching for critical points for memory reasons. The density fields used in this Section produced with 1LPT/1PPT with CDM and FDM initial conditions.

\subsubsection{Total number of critical points}

\begin{figure}[h!t]
\centering\includegraphics[width=\columnwidth]{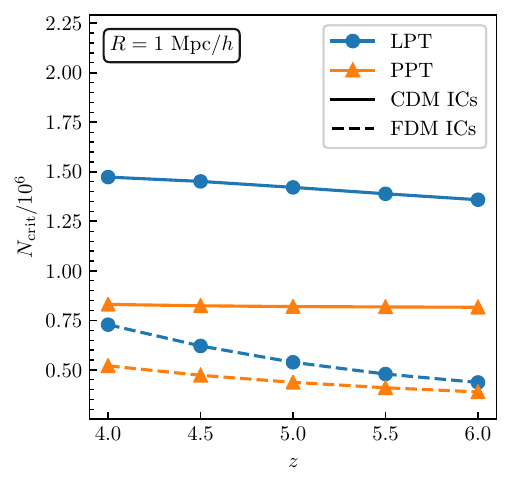}
    \caption{Total number of critical points on the density field smoothed on $R=1 \ \mathrm{Mpc}/h$ as a function of redshift.}
    \label{fig:crit_points_Ntot}
\end{figure}

\begin{figure}[h!t]
\centering\includegraphics[width=\columnwidth]{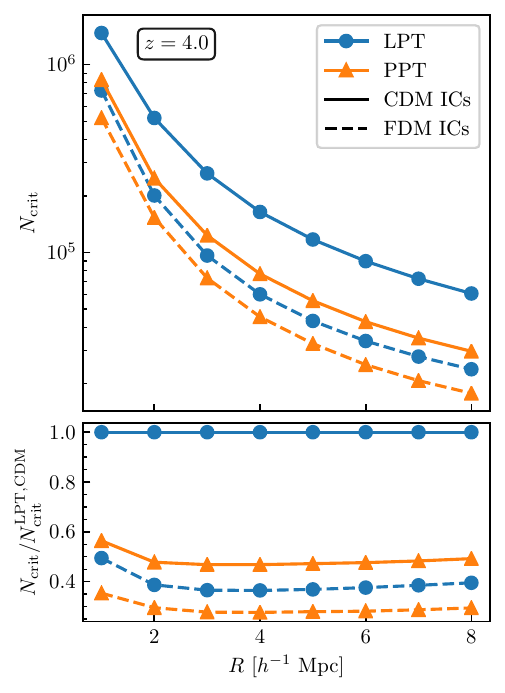}
    \caption{Total number of critical points at $z=4$ as a function of top-hat smoothing scale $R$. The lower panel shows the ratio to the number of critical points in the LPT + CDM ICs simulation, which is roughly constant on scales larger than $1 \ h^{-1} \ \rm Mpc$. The error on the mean based on 8 subboxes is smaller than the marker size in all panels.}
    \label{fig:crit_points_Ntot_vs_R}
\end{figure}

Figure~\ref{fig:crit_points_Ntot} shows the evolution of the total number of critical points over time, Figure~ \ref{fig:crit_points_Ntot_vs_R} shows the total number of critical points on different smoothing scales at redshift $z=4$. We see that at all redshifts, any amount of smoothing in the density field, whether from initial conditions or introducing wave dynamics decreases the total number of critical points. This is mirrored as the smoothing scale is increased, which also decreases the total number of critical points. However for radii larger than about $1 \ h^{-1} \ \rm Mpc$, the ratio of number of critical points in the different simulation runs remains roughly constant. We additionally see a stronger redshift evolution of the number of critical points in classical dynamics compared to wave dynamics. Higher amounts of smoothing decrease the total number of critical points, however the relative fraction of different critical points do not all decrease at the same rate, as we see reflected in the ratios and number fractions between different types of critical points in Figures \ref{fig:crit_points_ratio_evolution} \& \ref{fig:crit_points_fraction_evolution}. 

We note as well that the effect of the dynamics (LPT vs PPT, blue vs orange) in this case is stronger than seen in the one-point statistics on the same scale. While the change to the initial conditions provides the bulk of the suppression in the number of critical points, there is a larger separation between the LPT+FDM ICs (blue dashed) and PPT+FDM ICs (orange dashed) cases than seen in the reduced skewness in Figure \ref{fig:S3_1column}, indicating that critical points are more sensitive to wave interference effects, as expected from a probe of derivatives of the density field.

\subsubsection{Fraction of cosmic web elements}

In a Gaussian random field the ratio of critical points of different kinds is exactly solvable. The number of peaks and voids (or filaments and walls) are equal, while the ratio of filaments to peaks (or walls to voids) is \citep{Bardeen.etal_1986_StatisticsPeaks}
\begin{subequations}
\begin{align}
    \frac{N^{\rm G}_{\mathcal{P}}}{N^{\rm G}_{\mathcal{V}}}&=\frac{N^{\rm G}_{\mathcal{F}}}{N^{\rm G}_{\mathcal{W}}}=1,  \\
    \frac{N^{\rm G}_{\mathcal{F}}}{N^{\rm G}_{\mathcal{P}}}&=\frac{N^{\rm G}_{\mathcal{W}}}{N^{\rm G}_{\mathcal{V}}}= \frac{29\sqrt{15}+18\sqrt{10}}{29\sqrt{15}-18\sqrt{10}}\,,
\end{align}
\end{subequations}
and therefore, the number fractions for a Gaussian random field $f_i^{\rm G}=N_{i}^{\rm G}/N_{\rm crit}^{\rm G}$ are
\begin{subequations}
\begin{align}
    f_\mathcal{P,V}^{\rm G} &=  \frac{29- 6\sqrt{6}}{116} \approx 12.3\%, \\
    f_\mathcal{F,W}^{\rm G} &= \frac{29+ 6\sqrt{6}}{116} \approx 37.7\%\,.
\end{align}
\end{subequations}

\begin{figure}[t]
    \centering
\includegraphics[width=\columnwidth]{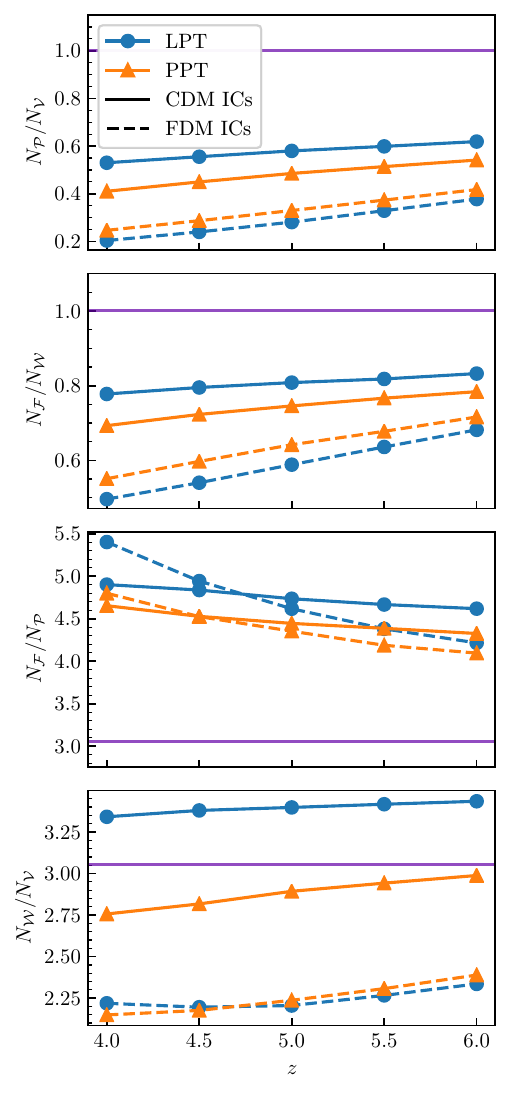}
    \caption[The ratios of critical points of different types.]{Ratio of number of critical points of different types. The horizontal solid purple lines show the ratios for a Gaussian random field. }
    \label{fig:crit_points_ratio_evolution}
\end{figure}

Figure~\ref{fig:crit_points_ratio_evolution} shows these ratios as measured on $R=1 \ h^{-1} \ \rm Mpc$ together with the Gaussian random field values. We see that generally the fully classical (LPT + CDM ICs) systems are the closer to the Gaussian ratios, with either FDM ICs or wave dynamics introducing higher deviation from the Gaussian ratios. This is consistent with the skewness results, which indicate that FDM initial conditions and wave dynamics both act to produce a more non-Gaussian field, though the dynamical effects have a stronger impact on the non-Gaussianity as measured from the critical points than the skewness.

\begin{figure}[t]
    \centering
\includegraphics[width=\columnwidth]{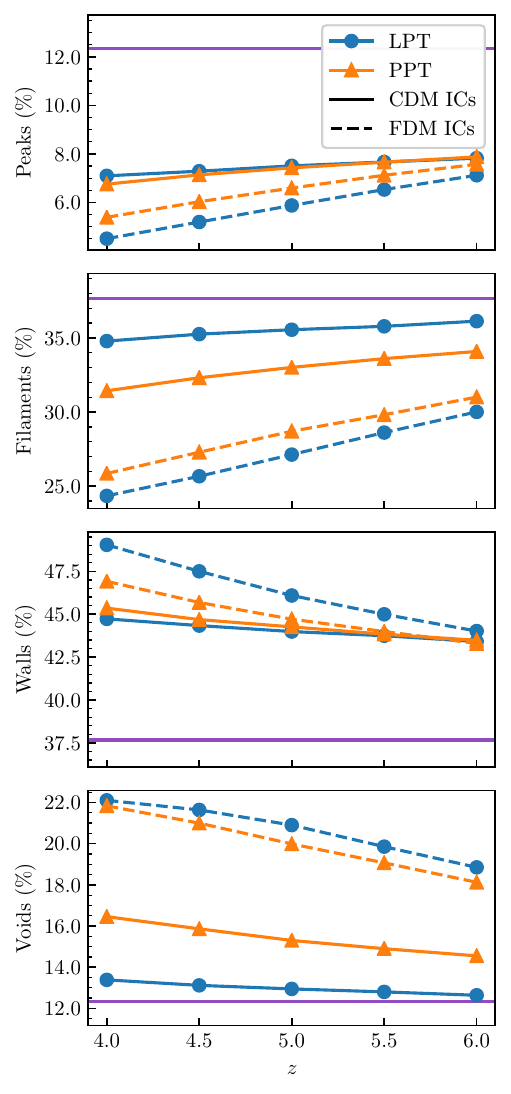}
    \caption[Evolution of the fraction of critical points of a given type.]{Evolution of the fraction of critical points on $R=1 \ h^{-1} \ \rm Mpc$ of a given type. The horizontal solid purple lines show the fraction for a Gaussian random field. Both wave dynamics (orange) and FDM initial conditions (dashed lines) move these fractions further from their Gaussian values. }
    \label{fig:crit_points_fraction_evolution}
\end{figure}
Figure~\ref{fig:crit_points_fraction_evolution} shows the evolution of the fraction critical points in 1 $h^{-1}$ Mpc spheres of different types across the different cosmologies. 

We see that at all redshifts and for all of the cosmologies, the underdense type critical points (voids and walls) are both enhanced from the fractions which would be expected in a Gaussian field ($\sim \!12\%$ for voids and $\sim\! 38\%$ for walls) with smoothed initial conditions particularly enhancing this enhancement. While we only show the results for 1LPT/1PPT in Figure \ref{fig:crit_points_fraction_evolution}, the same behaviour is seen in the $2^{\rm nd}$ order perturbative results as well.

From these individual number fractions, we see that while the FDM ICs or PPT dynamics cases produce fewer critical points than the fully classical case, as seen in Figures \ref{fig:crit_points_Ntot} and \ref{fig:crit_points_Ntot_vs_R}, the decrease is not uniform across different types of critical points. We see from Figure \ref{fig:crit_points_fraction_evolution} that underdense type critical points (voids and walls) are pushed further from their Gaussian predicted values than filaments and peaks. This can be explained by two effects. The removal of power on small scales in the FDM ICs cases reduces the gravitational potential, making it more difficult for structures to collapse in the first place, preferentially making void and wall type critical points more abundant over filaments and peaks. Secondly, the quantum pressure term in the wave dynamics case (equation~\ref{eqn:PPT-continuity}) acts to reduce shell crossing compared to the LPT dynamics case, which reduces the overdense type critical points more.

We note that \cite{Dome.etal_2023_CosmicWebDissection} examined the mass and volume fractions of different cosmic web elements in the context of classical FDM (analagous to our LPT + FDM ICs case) and found that their volume fraction of voids decreased as the initial conditions were suppressed, contrary to our findings here. However, their cosmic web classification uses the NEXUS+ algorithm to classify all points into a cosmic web element, based on the signs of the Hessian over a range of scales, rather than critical points specifically, so direct comparison of these results is difficult. Much of their analysis also takes place on significantly smaller scales ($\sim \! 39 \ h^{-1}\  \rm kpc$, their grid scale) than we consider, where more non-linear dynamics is relevant.

\subsubsection{Environment spilt PDFs}

\begin{figure}[h!t]
\centering
\includegraphics[width=\columnwidth]{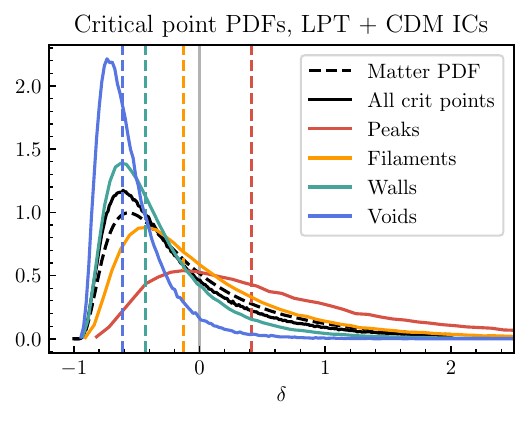}
\caption[The overdensity PDFs in different critical point environments.]{The PDFs of the overdensity of different critical points of the fully classical (LPT + CDM initial conditions) density field at $z=4$ when smoothed on $R=1$ Mpc/$h$. Each individual PDF is normalised. The PDF of the entire matter field (black dashed), and of all critical points (black solid) are shown for comparison. Coloured vertical lines show the median overdensity for the critical points of that type. The grey vertical line divides over- and underdense regions.}
\label{fig:crit_point_env_split_pdfs}
\end{figure}

\begin{figure*}[h!t]
    \centering
\includegraphics[width=2\columnwidth]{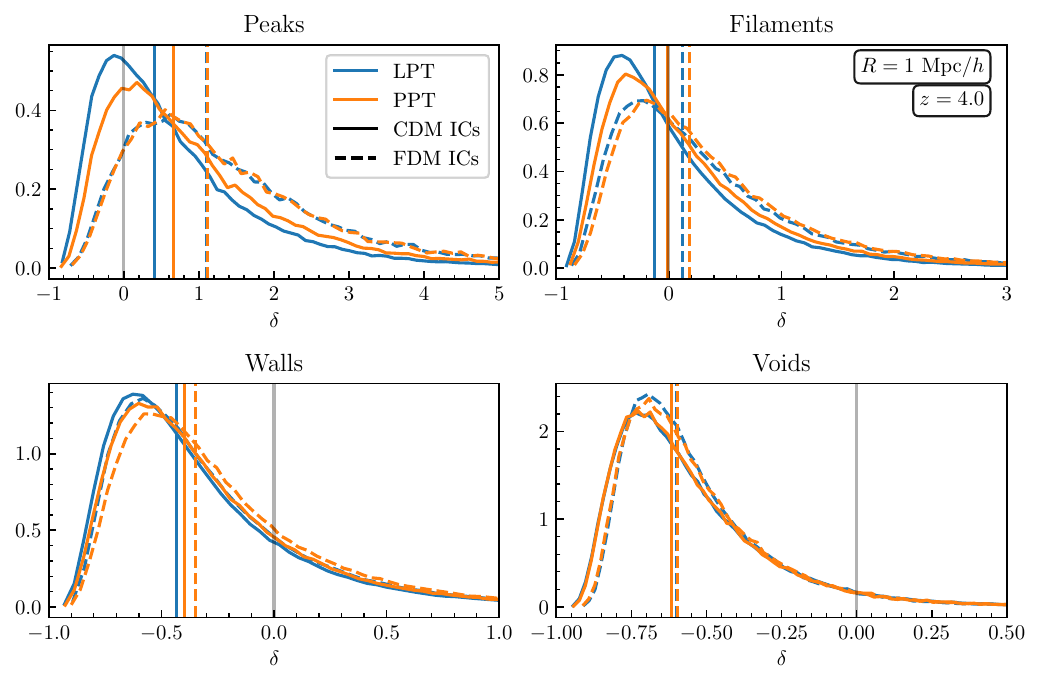}
    \caption[The overdensity PDFs in different environments for different initial conditions and dynamics.]{The PDFs of the density found in different critical point environments, linearly interpolated between bins. The vertical lines show the median value of $\delta$ for that cosmology, the grey vertical line divides under- and over-dense environments. Environments with more shell crossing, such as peaks, have a larger shift when wave dynamics are turned on, due to the quantum potential having a larger effect. }
    \label{fig:crit_point_PDFs_2x2}
\end{figure*}

In addition to the number of critical points of certain types we can also examine the mass distribution of different critical points. Figure \ref{fig:crit_point_env_split_pdfs} shows the  overdensity PDFs for different critical point environments $z=4$ on $R=1 \ h^{-1} \ \rm Mpc$ scales in the fully classical (LPT + CDM initial conditions) density field. 
We see that as expected, peaks are the densest critical points, while voids are the most underdense. We see the variation of these PDFs across cosmologies in Figure~\ref{fig:crit_point_PDFs_2x2}. We see that wave dynamics and FDM ICs both increase the density of peaks, shifting the PDFs to the right. Interestingly, while the number of voids was heavily impacted by introducing FDM ICs or wave dynamics, as seen in Figure~\ref{fig:crit_points_fraction_evolution}, the void PDF is relatively stable compared to the other environment changes. These effects can be understood by the effect of the quantum potential,
\begin{equation}
Q = - \frac{\hbar_{\rm PPT}^2}{2}\frac{\nabla^2 \sqrt{1+\delta}}{\sqrt{1+\delta}},
\end{equation}
as it is dependent on the curvature of the (square root) of the density, and therefore is largest in regions of high curvature. We see that for cold initial conditions (solid lines) the critical points involving more shell crossing have larger shifts to their PDFs due to this quantum pressure. However, on FDM initial conditions, the PDFs of critical point environments seem less sensitive to the role of this quantum potential, even in the peaks where it was previously the strongest. We expect this is due to the suppressed initial conditions removing the critical points on the smallest scales, which also correspond to the highest curvature environments where quantum pressure would be the strongest.

The behaviour of the PDFs we see on these scales is quite different to those found in \citet{Dome.etal_2023_CosmicWebDissection} in the context of classical FDM, where they find peaks to have the most stable PDF across changing cosmologies, and their voids have the most sensitivity to changing cosmology. However this is not immediately surprising as their PDFs are extracted on much smaller scales ($0.04$ vs $1$ Mpc/$h$), and their environments are quantified differently (for example, their void PDFs quantify all the matter in the void, while ours consider the PDF of densities at the void-type critical points, roughly the density at the centre of voids). While the behaviour between different environments is quite different, the overall effect of shifting all PDFs towards more overdense values is seen in both cases.

We note as well that in the fully non-linear case, the PDF of peak type critical points is likely to look different on small scales, as in \citet{Dome.etal_2023_CosmicWebDissection}, as a consequence of virialisation creating stable halo profiles which feed into the peak PDF. Since our work considers 1PPT, equivalent to the Zel'dovich approximation in the $\hbar_{\rm PPT}\to 0$ limit, our model will not produce virialised objects due to overshooting, and thus we do not see the stability of the peak PDF discussed in \citet{Dome.etal_2023_CosmicWebDissection}.

\section{Conclusions}

\textit{Summary.}
In this paper we present an analysis of density fields produced by  two perturbative forward models on scales currently inaccessible to full Schr\"odinger-Poisson solvers. The models considered encode either classical fluid dynamics (LPT) or wave dynamics (PPT). Both dynamics were applied to both CDM and FDM initial conditions to disentangle effects of initial conditions from dynamical interference effects. We demonstrate that the wave-perturbation scheme used provides similar effects on the power spectrum to fully non-linear Schr\"odinger-Poisson simulations.

For statistics of density environments such as the matter PDF  or the reduced skewness $S_3$ on mildly non-linear scales, we show that the principle difference between fully classical (LPT + CDM ICs) and fully wave (PPT + FDM ICs) dark matter is driven mostly by the suppression of the initial conditions. This gives credibility to the ``classical fuzzy dark matter'' \citep{Dome.etal_2023_CosmicWebDissection} approach on large perturbative scales for these sorts of averaged cell statistics.

Analysis of the critical points of the density field show that these density field extrema are more sensitive to interference and quantum pressure effects than averaged one-point statistics. This is expected as both the quantum potential and these extrema are sensitive to derivatives of the underlying density field. Both suppression of the initial power and wave interference affect the total number and relative fraction of critical points, decreasing the amount of gravitational collapse and shell crossing, preferentially erasing critical points which require more shell crossing (peaks and filaments) compared to underdense type critical points (voids and walls). The one-point statistics of these different critical point environments also clearly demonstrate the role of the quantum pressure, with more collapsed regions producing a larger shift in the PDF compared to the fully classical case.

\textit{Outlook.}
Our analysis suggests that the one-point PDF and skewness of the density field appear fairly insensitive to the presence of wave dynamics on mildly nonlinear scales. The prospect for theoretically modelling such statistics in fully non-linear FDM cosmologies is therefore encouraging, requiring only changes to the linear power spectrum. The large-deviations theory model of the matter PDF \citep{Bernardeau14,Uhlemann16} has successfully been adapted to non-$\Lambda$CDM cosmologies \citep{Uhlemann:2020, Cataneo.etal_2022_MatterDensity}. Because of the promise of the PDF as a non-Gaussian probe of extended cosmologies, a similar forecast or inference based on the matter PDF and related quantities could be a powerful complement to existing statistics in detecting wavelike dark matter effects. Successful modelling of the dark matter PDF in FDM cosmologies could then be extended to observables such as biased tracers \citep{Uhlemann:2018b, Friedrich_2021} or weak lensing statistics \citep{Reimberg2018PhRvD, Uhlemann:2018a, Barthelemy:2020,Barthelemy2021MNRAS, Barthelemy2022PhRvD, Barthelemy2023arXiv, Boyle.etal_2023_CumulantGenerating}, as has been done in $\Lambda$CDM cosmologies. 

A variety of other statistics beyond those presented in this work could be extracted from this map-based forward model, further pushing the question of whether the quantum potential imprints signatures on any large scale statistics which are not captured by suppressing the initial conditions. Statistics related to the velocity and velocity dispersion could be extracted by utilising the phase information of the PPT wavefunction. Spatial correlations of/between the critical points of the density field could provide additional information to simple number counts and one-point statistics presented here. Additionally, analysing the sensitivity of different cosmic web identification methods would be a natural extension to this work and could provide insights into the structure of the cosmic web in fuzzy cosmologies.

Our wave-mechanical models present an appealing complementary approach to large scale studies of fuzzy dark matter. With further theoretical work to establish a mapping between $\hbar_{\rm PPT}$ and the FDM particle mass $m$, propagator perturbation theory could be used as a less numerically intensive method of generating simulations of wave dark matter on large scales. While current hybrid approaches patch large scale $N$-body with small scale Schr\"odinger-Poisson solvers, it could be possible to patch this wave based perturbation theory on large scales with Schr\"odinger-Poisson on small scales, in a similar spirit to \texttt{COLA} \citep{Tassev.etal_2013_SolvingLarge} and \texttt{Hi-COLA} \citep{Wright.etal_2023_HiCOLAFast}. Large scale simulations of wave dark matter which maintain interference effects on all scales would be an important tool for testing if the fundamental nature of dark matter is particle-like or wave-like and derive constraints from astrophysical data.

\section*{Acknowledgements} 

We thank the anonymous referees for valuable suggestions that improved the presentation of our
results. We thank Oliver Hahn, Cornelius Rampf, and Mateja Gosenca for useful conversations and feedback at multiple stages of this project. AG thanks Alexandre Barthelemy and Oliver Friedrich for discussions with initialised some of this project's ideas. We thank Simon May for discussions and sharing the power spectra data.

The figures in this work were created with \textsc{matplotlib} \citep{matplotlib} making use of the \textsc{numpy} \citep{numpy}, \textsc{scipy} \citep{2020SciPy-NMeth}, and \textsc{scikit-image} \citep{scikit-image} Python libraries. 

AG was supported by an EPSRC studentship under Project 2441314 from UK Research \& Innovation. CU's research was supported in part by grant NSF PHY-1748958 to the Kavli Institute for Theoretical Physics (KITP) and in part funded by the European Union (ERC StG, LSS\_BeyondAverage, 101075919). AG and CU are grateful for the hospitality of Perimeter Institute where this work was finalised following peer review. Research
at Perimeter Institute is supported in part by the Government of Canada through the Department of
Innovation, Science and Economic Development and by the Province of Ontario through the Ministry of
Colleges and Universities. This research was also supported in part by the Simons Foundation through the
Simons Foundation Emmy Noether Fellows Program at Perimeter Institute.

\section*{Data access statement}

The data in this paper was created using publicly available codes, using input parameters detailed in the body of the text. For the purpose of open access, the authors have applied a `Creative Commons Attribution' (CC BY) licence to this paper.

\bibliographystyle{aasjournal}
\bibliography{refs}

$\,$
\appendix

\section{Matching PPT scale and FDM mass}\label{app:sec:matching_PPT_to_LPT}
\subsection{Effect of larger $\hbar_{\rm PPT}$}\label{app:sec:largerhbar}

\begin{figure}[b]
    \centering
    \includegraphics[width=\columnwidth]{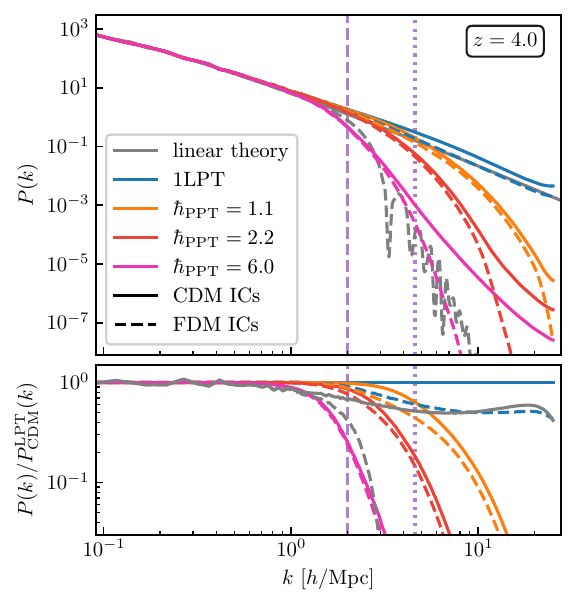}
    \caption{Same as Figure \ref{fig:Pk-z4} with two additional values of $\hbar_{\rm PPT}$. The power spectrum with $\hbar_{\rm PPT}=6 \ h^{-2} \ \rm Mpc^2$ sees comparable suppression to the linear FDM power spectrum (grey dashed) at $z=4$. Full Schr\"odinger-Poisson simulations indicate these scales should have been enhanced by non-linear evolution by redshift 4, indicating that $\hbar_{\rm PPT}=6 \ h^{-2} \ \rm Mpc^2$ is not well matched to the mass $m_{22}=0.1$.}
    \label{fig:pk_2hbars}
\end{figure}

\begin{figure}
    \centering
    \includegraphics[width=\columnwidth]{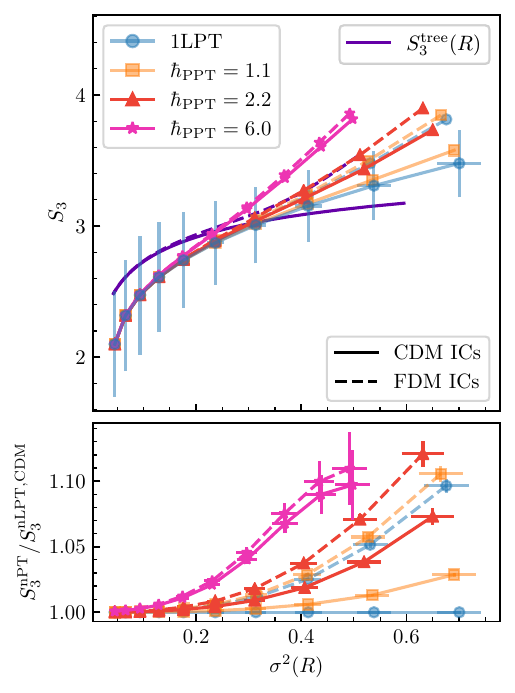}
    \caption{Same as Figure~\ref{fig:S3_1column} with the additional values $\hbar_{\rm PPT}$. The larger value of $\hbar_{\rm PPT}=2.2 \ h^{-2} \ \rm Mpc^{2}$ is able to produce the skewness of the LPT + FDM ICs case even on cold initial conditions. The $\hbar_{\rm PPT}=6 \ h^{-2} \ \rm Mpc^{2}$ produces a significantly enhanced $S_3$, driven by the dynamics rather than the initial conditions. The lines previously shown are made slightly transparent to aid readability.}
    \label{fig:s3_2hbars}
\end{figure}

Figures \ref{fig:pk_2hbars} and \ref{fig:s3_2hbars} show the power spectrum and reduced skewness as in Figures \ref{fig:Pk-z4} and \ref{fig:S3_1column}, now with an additional values of $\hbar_{\rm PPT}$. A value of $\hbar_{\rm PPT} = 2.2 \, h^{-2} \ \rm Mpc^2$ (twice the value used in the main analysis) produces skewness enhancements similar to those due to changing the initial conditions through wave dynamics alone (compare the red/triangle solid line to the blue/circle or orange/square dashed lines). The choice of $\hbar_{\rm PPT} = 6 \, h^{-2} \ \rm Mpc^2$ leads to a suppression in the power spectrum similar to the linearly evolved FDM power spectrum for $m_{22}=0.1$. Since full Schr\"odinger-Poisson simulations indicate that these scales should see power exceeding the linearly predicted value (see Figure \ref{fig:Pk-ratios-with-data}) this suggests that this choice of $\hbar_{\rm PPT}$ is not well matched to the mass used to set the initial conditions $m_{22}=0.1$.  Based on this we do not expect $m_{22}=0.1$ to map to an $\hbar_{\rm PPT}$ greater than 6 $h^{-2}$ Mpc$^2$. In this sense, our choice of $\hbar_{\rm PPT}=1.1 \ h^{-2} \ \rm Mpc^2$ provides a conservative estimate of the effect of wave dynamics compared to initial conditions. It may be possible to determine a more precise mapping between $\hbar_{\rm PPT}$ and $m_{22}$ for which ideas are sketched in Appendix~\ref{app:mappinghbar}.

\subsection{The role of time variables}
Here we comment on the difference in coordinates between the PPT wave dynamics and the true FDM style equations (which are the same as the Widrow-Kaiser style application of Schr\"odinger-Poisson to modelling CDM). This is also discussed in \citet{Uhlemann2019}.

The PPT equations  take the scale factor $a$ as the time variable, rather than the cosmic time $t$ as in equations \eqref{eqn:SP-full-eqns}. This impacts the natural velocity/momentum variable in the two pictures: in PPT it is $\bm{v} = \dv*{\xx}{a} = \grad \phi_v$ which differs from the  peculiar velocity $\bm{u} = \dv*{\xx}{t} = \dot{a}a\bm{v} = Ha^2\bm{v}$ which is generated by the wavefunction phase in the FDM scenario \eqref{eqn:SP-fluid-eqns:Bernoulli}. This leads to the identification $\phi = \dot{a} a^2 \phi_v$ as the mapping between the phase in the FDM picture and the PPT picture. These variable changes amount to a non-canonical transformation in phase space, and are what lead to the differing powers of $a$ in the quantum potentials in the two formulations. Specialising to an Einstein-de Sitter universe, where $\dot{a} \overset{\rm EdS}{=} H_0 a^{-1/2}$ and $\Omega_m^0 = 1$, the relationship between these wavefunction phases is  $\phi  \overset{\rm EdS}{=}H_0 a^{3/2}\phi_v$. 

Rewriting the fluid equation \eqref{eqn:SP-fluid-eqns} for FDM in $a$-time, we can make more direct comparison between these systems. The continuity equation becomes the same as that in PPT \eqref{eqn:PPT-continuity}, under the mapping between phase variables. Defining the rescaled the gravitational potential by $\varphi_g \overset{\rm EdS}{=} 2V_N/(3H_0^2)$, the FDM Poisson equation \eqref{eqn:SP-fluid-equns:Poisson} becomes the PPT Poisson equation \eqref{eqn:PPT-poisson-fluid}. The Bernoulli equation \eqref{eqn:SP-fluid-eqns:Bernoulli} becomes 
\begin{align}
\del_a \phi_v + \frac{1}{2}\abs{\nabla\phi_v}^2 
&\overset{\mathrm{EdS}}{=} -\frac{3}{2a}(\varphi_g + \phi_v) + \frac{\hbar^2}{2a^3H_0^2} \frac{\nabla^2\sqrt{1+\delta}}{\sqrt{1+\delta}} .
\label{eqn:SP-fluid-eqns:Bernoulli_ppt}
\end{align}
Comparing this to the PPT Bernoulli equation \eqref{eqn:PPT-continuity} we see that the quantum potential term carries an additional factor of $a^3$ in EdS spacetimes (more generally a factor of $(\dot{a}a^2)^{2}$), leaving the $\hbar_{\rm PPT}$ to be interpreted as a time dependent coarse-graining scale, or because the quantum potential in the true FDM case is controlled by the boson mass, $\hbar_{\rm PPT}$ acts like a time dependent mass.

\subsection{Mapping $\hbar_{\rm PPT}$ to FDM mass}
\label{app:mappinghbar}
As the PPT velocity potential $\phi_v$ generates the velocity $\bm{v}=\dv*{\xx}{a}$ which has dimensions of length, $\hbar_{\rm PPT}$ must then have dimensions of $[\mathrm{length}]^2$. The mapping between between the semiclassical parameter $\hbar_{\rm PPT}$ and the physical FDM particle mass $m_{22}$ isn't straightforwardly 1-to-1 due to the different time evolutions of the quantum potential terms (effectively the $\hbar_{\rm PPT}$ time dependence). Additionally, numerical applications of these systems require different initial conditions, with true FDM requiring initial conditions set by the physical dynamics at some fixed early time, while perturbation theories formally evolve from an initial time of $a\to 0$ on tethered initial conditions like those discussed in equation \eqref{eqn:ZA_boundary_conditions}.

For the purposes of this investigation we take a code-driven approach to choosing $\hbar_{\rm PPT}$ associated with a mass $m_{22}$, which we discuss in Section~\ref{sec:howw-fuzzy-numerics}. Here we outline ideas for an analytic matching which could be implemented in future work.

The na\"ive approach to matching the FDM particle mass and the $\hbar_{\rm PPT}$ value is to simply equate the strength of the quantum potential terms in equations \eqref{eqn:SP-fluid-eqns:Bernoulli_ppt} and \eqref{eqn:PPT-Bernoulli} at some matching redshift $z_m$. For EdS, this results in 
\begin{equation}\label{eq:mass_hbar_mapping}
     \frac{\hbar_{\rm PPT} }{(\mathrm{Mpc}/h)^2} = \frac{1.34\times 10^{-4}}{m_{22}} (1+z_m)^{-3/2} \left(\frac{h}{0.7}\right).
\end{equation}
If we fix $m_{22}=0.1$ as the smallest value of FDM particle mass of interest at  $z_m=10$, the resulting $\hbar_{\rm PPT}\sim$ 0.05 $h^{-2} \rm \ Mpc^2$, which is too small to resolve on a reasonably computationally cheap grid scale without aliasing in the wavefunction. When choosing a matching redshift of $z_m=100$, one gets $\hbar_{\rm PPT}\sim 1.4 \ h^{-2} \rm \ Mpc^2$, which is close to our choice. However, the matching prescription \eqref{eq:mass_hbar_mapping} is crude as the physical effect of interest is the integrated impact of the quantum potential term on the density field. Neglecting the spatial variation due to density (and $\delta\rightarrow 0$ as $a\rightarrow 0$), the matching cannot be extended all the way to $a_{\rm s}\to 0$ as the perturbation theory would require. However, as the quantum potential should only be relevant once enough curvature in the density field is sourced---which will later cause interference in regions of classical shell crossing---this integration should only be carried back to the onset of structure formation, expected to be relevant from around $z_s=\mathcal O(10-100)$. 
If we assumed the Schr\"odinger-Poisson quantum potential in equation~\eqref{eqn:SP-fluid-eqns:Bernoulli_ppt} scales as $Q_{m_{22}}(a)$ compared to a PPT quantum potential in equation~\eqref{eqn:PPT-Bernoulli} scaling like $Q_{\hbar_{\rm PPT}}(a)$, matching their integrated effects aggregated over a period from $a_{\rm s}$ to $a_{\rm f}$ corresponds to computing a correction factor like
\begin{equation}
    f_{\hbar_{\rm PPT},\rm eff}(z_{\rm s},z_{\rm f}) \propto \sqrt{\frac{\int_{a_{\rm s}}^{a_{\rm f}} da\, Q_{m_{22}}(a)}{\int_{a_{\rm s}}^{a_{\rm f}} da \,Q_{\hbar_{\rm PPT}}(a)}}\,,
\end{equation}
where $a=(1+z)^{-1}$. Neglecting the temporal variation of the density, one might assume $Q_{m_{22}}(a)=a^{-3}$ and $Q_{\hbar_{\rm PPT}}(a)=1$. This leads to $f_{\hbar_{\rm PPT},\rm eff}(4,10)=21$ and $f_{\hbar_{\rm PPT},\rm eff}(4,100)=164$. When considering the leading order temporal evolution of $\delta\propto a$, we would have $Q_{m_{22}}=a^{-2}$ and $Q_{\hbar_{\rm PPT}}(a)=a$, which leads to values of the same order of magnitude $f_{\hbar_{\rm PPT},\rm eff}(4,10)=19$ and $f_{\hbar_{\rm PPT},\rm eff}(4,100)=70$, and hence a similar order of magnitude.

A more precise analysis of what redshift shell crossing becomes sufficiently large to affect this could for example be investigated in a simple one-dimensional comparison between PPT and a full Schr\"odinger-Poisson solver. 

\begin{figure}[b]
\centering
    \includegraphics[width=\columnwidth]{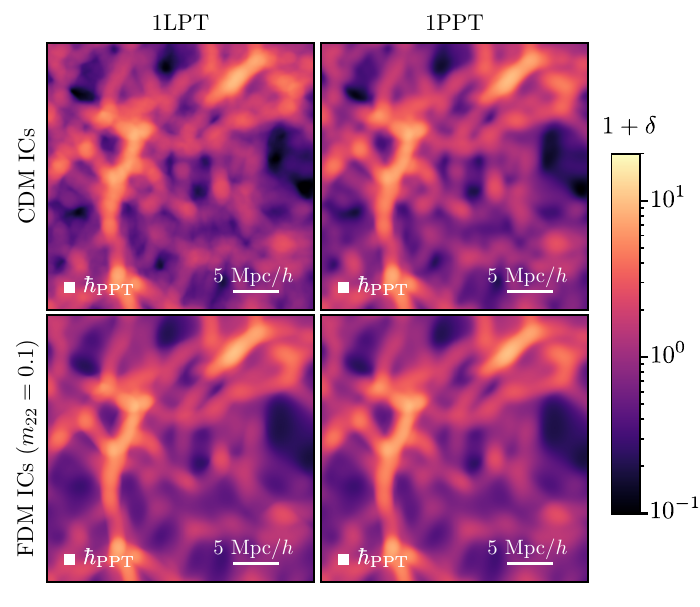}
    \caption{Projected 0.125 Mpc/$h$ slice of the dark matter density at $z=4.0$, as in Figure \ref{fig:slices-of-log-delta}, after the fields have been smoothed with a top-hat filter of radius $R=1$ Mpc/$h$. These show the same region as the zoomed insets in Figure \ref{fig:slices-of-log-delta} and are shown on the same colour scale.}
    \label{fig:density_smoothed}
\end{figure}

\begin{figure*}
    \centering
    \includegraphics[scale=1]{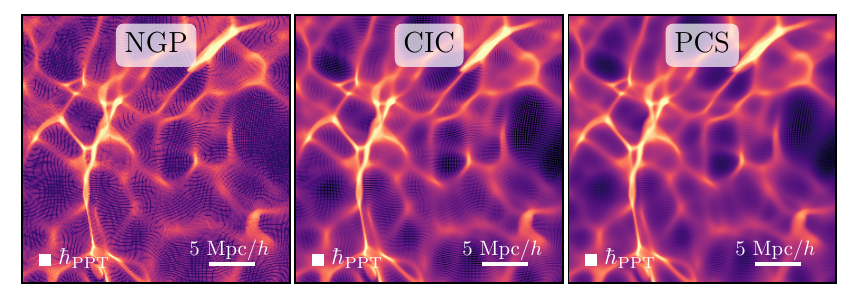}
    \caption{Density fields for 1LPT + FDM ICs as in Figure \ref{fig:slices-of-log-delta}, where particle positions have been deposited on the grid by different mass assignment schemes: nearest grid point (NGP), cloud-in-cell (CIC), piecewise cubic spline (PCS). The fine grained crosshairs seen in the underdense regions disappear as the mass assignment order increases.}
    \label{fig:MAS_comparison}
\end{figure*}

\section{Impact of mass assignment scheme}\label{app:sec:MAS_impact}

Figure \ref{fig:MAS_comparison} shows the density field constructed from particle positions for 1LPT + FDM ICs using different mass assignment schemes of increasing order. The fine grained pattern, particularly noticible in the underdense region worsens with lower order mass assignment schemes such as nearest grid point (NGP) and smooths out with higher order schemes such as piecewise cubic spline (PCS). For statistics in the main analysis we analyse the smoothed density field, such as that shown in Figure \ref{fig:density_smoothed}, where these fine grained patterns disappear in cloud-in-cell (CIC) assignment. We make use of CIC mass assignment to construct density fields from LPT displacements throughout the rest of this work.

\end{document}